\documentclass{ieeeaccess}
\usepackage{cite}
\usepackage{amsmath,amssymb,amsfonts}
\usepackage{graphicx}
\usepackage{textcomp}

\usepackage{bm}
\makeatletter
\AtBeginDocument{\DeclareMathVersion{bold}
\SetSymbolFont{operators}{bold}{T1}{times}{b}{n}
\SetSymbolFont{NewLetters}{bold}{T1}{times}{b}{it}
\SetMathAlphabet{\mathrm}{bold}{T1}{times}{b}{n}
\SetMathAlphabet{\mathit}{bold}{T1}{times}{b}{it}
\SetMathAlphabet{\mathbf}{bold}{T1}{times}{b}{n}
\SetMathAlphabet{\mathtt}{bold}{OT1}{pcr}{b}{n}
\SetSymbolFont{symbols}{bold}{OMS}{cmsy}{b}{n}
\renewcommand\boldmath{\@nomath\boldmath\mathversion{bold}}}
\makeatother

\def\BibTeX{{\rm B\kern-.05em{\sc i\kern-.025em b}\kern-.08em
    T\kern-.1667em\lower.7ex\hbox{E}\kern-.125emX}}

\usepackage{float}
\usepackage{adjustbox}
\usepackage{placeins}
\usepackage[all]{xy}
\usepackage{enumitem}
\usepackage{mathtools,amssymb}
\usepackage{algorithm}
\usepackage[noend]{algpseudocode}
\usepackage{csquotes}

\renewcommand{\it}{\itshape}
\renewcommand{\sl}{\slshape}    
\renewcommand{\sc}{\scshape}
     
\renewcommand{\bf}{\bfseries}

\newcommand{\Nbb}{\mathbb{N}}
\newcommand{\Qbb}{\mathbb{Q}}  
\newcommand{\Rbb}{\mathbb{R}}

\newcommand{\Dcal}{\mathcal{D}}
\newcommand{\Mcal}{\mathcal{M}}  
\newcommand{\Ncal}{\mathcal{N}}

\newcommand{\q}{\quad}              
\newcommand{\qq}{\qquad}

\newcommand{\ra}{\rightarrow}       
        
\newcommand{\Lra}{\Leftrightarrow}

\renewcommand{\epsilon}{\varepsilon}    
\renewcommand{\phi}{\varphi}
            
\newcommand{\E}{\operatorname{E}}
\newcommand{\Var}{\operatorname{Var}}

\DeclareMathOperator*{\argmax}{arg\,max}

\begin{document}

\history{Date of publication xxxx 00, 0000, date of current version xxxx 00, 0000.}
\doi{10.1109/ACCESS.2023.1120000}

\title{A Gaussian process based approach for validation of
multi-variable measurement systems: application to SAR measurement systems}

\author{\uppercase{Cédric Bujard}\authorrefmark{1}, 
\uppercase{Esra Neufeld}\authorrefmark{1}, 
\uppercase{Mark Douglas}\authorrefmark{1},
\uppercase{Joe Wiart}\authorrefmark{2}, 
\uppercase{Niels Kuster}\authorrefmark{1, 3}}

\address[1]{IT'IS Foundation, Zurich, Switzerland}
\address[2]{Laboratoire de Traitement et Communication de l’Information (LTCI), C2M, Télécom Paris, Paris, France}
\address[3]{ETH, Zurich, Switzerland}

\corresp{Corresponding authors: Cédric Bujard (e-mail: bujard@itis.swiss), 
Mark Douglas (e-mail: douglas@itis.swiss).}

\begin{abstract}
Resource-efficient and robust validation of systems designed to measure a
multi-dimensional parameter space is an unsolved problem as it would require
millions of test permutations for comprehensive validation coverage. In the
paper, an efficient and comprehensive validation approach based on a
Gaussian Process (GP) model of the test system has been developed that can
operate system-agnostically, avoids calibration to a fixed set of known
validation benchmarks, and supports large configuration spaces. The approach
consists of three steps that can be performed independently by different
parties: 1) GP model creation, 2) model confirmation, and 3) targeted search
for critical cases. It has been applied to two systems that measure specific
absorption rate (SAR) for compliance testing of wireless devices and apply
different SAR measurement methods: a probe-scanning system (per IEC/IEEE
62209-1528), and a static sensor-array system (per IEC 62209-3). The results
demonstrate that the approach is practical, feasible, suitable for proving
effective equivalence, and can be applied to any measurement method and
implementation. The presented method is sufficiently general to be of value
not only for SAR system validation, but also in a wide variety of
applications that require critical, independent, and efficient validation.
\end{abstract}


\titlepgskip=-21pt

\maketitle

\section{Introduction} 

The operation of wireless devices close to the body results in millions of
distinct induced electromagnetic (EM) field distributions. Accurate assessment
of these fields is a complex task for which different measurement approaches
have been proposed and defined in two different compliance testing standards. In
this study, we have derived the key requirements to ensure reliable,
independently-verifiable, implementation-agnostic, and comprehensive validation.
For that purpose, a general, Gaussian Process (GP) model-based approach has been
developed and is under discussion for adoption in upcoming revisions of the
specific absorption rate (SAR) measurement standards \cite{iec62209-1528,
iec62209-3}. The data and the source code of an implementation of the procedure
are available at \cite{bujard}, and a corresponding application is accessible
through a graphical user interface (GUI) at \cite{bujard2}. The approach, which
is elaborated and demonstrated in this paper, is believed to be widely
applicable, i.e., well beyond SAR system validation, in situations where similar
validation requirements exist.

\subsection{SAR Measurement Systems}

SAR measurement systems measure the induced electric field distribution in
phantoms that represent the user of a commercial wireless device. That wireless
device is operated in close proximity to the phantom, which contains a
tissue-simulating medium. At any location in the phantom, the SAR is related to
the root-mean-squared value of the induced E-field, $\vec{E}$, in the medium
through the relation 
\begin{align}
\mathrm{SAR} = {\frac{\sigma \lvert \vec{E} \rvert ^2}{\rho}},
\end{align}
where $\sigma$ is the frequency-dependent electrical conductivity and $\rho$ =
1000\,kg/m$^3$ is the mass density of the medium. The induced E-field is
temporally and spatially assessed in the phantom. Then, spatial and time
averaging of the SAR is applied, followed by the identification of the peak
spatial- and time-averaged SAR (psSAR). This value is compared against the psSAR
limits set by safety standards such as the International Commission on
Non-Ionizing Radiation (ICNIRP), which have been adopted by many regulators.
ICNIRP limits are also endorsed by the European Council \cite{ecRF} and adopted
in the harmonized standards of the European Committee for Electrotechnical
Standardization (CENELEC)~\cite{cenelec50360-2017, cenelec50566-2017}. Before a
wireless device can be sold on the market, the manufacturer is required to
demonstrate that the psSAR is below the limit in all tested conditions defined
by the measurement standard. ICNIRP has set a psSAR limit of $\text{SAR}_{10g}$
= 2~W/kg averaged over a 10\,g cubic mass, applicable to the general public for
localized exposures of the head and torso at frequencies from 100~kHz to
6~GHz~\cite{icnirp}. Some countries, such as Canada, India and USA, have adopted
a more stringent psSAR limit of $\text{SAR}_{1g}$ = 1.6~W/kg averaged over a
1\,g cubic mass in accordance with the IEEE Standard C95.1-1999~\cite{ieeec951}.
Other limits apply to the limbs, and there are different limits for occupational
exposure and whole-body exposure.

\begin{figure}[!ht] \centering
    \includegraphics[width=\columnwidth]{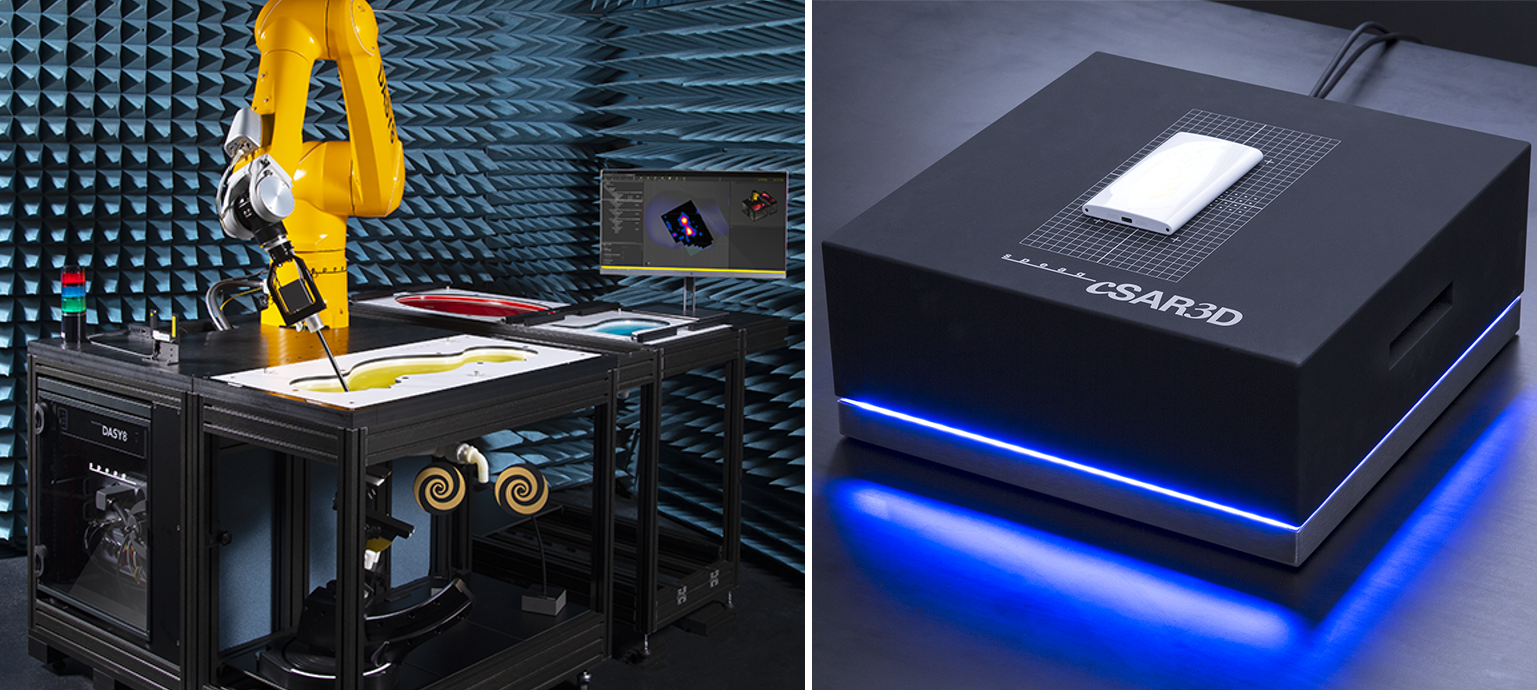}
    \caption{SAR measurement systems: DASY scanning system (left) and cSAR3D array system (right).}
    \label{fig:sarsystems}
\end{figure}

The two commonly used SAR measurement systems are scanning systems and array
systems.  Both systems will be studied in this paper using commercially
available products: the scanning system DASY8~\cite{pokovic2000} and the array
system cSAR3D~\cite{csar3d2011}, both manufactured by Schmid \& Partner
Engineering AG (Zurich, Switzerland) as shown in Fig~\ref{fig:sarsystems}.

Scanning systems use a robot to mechanically scan a single isotropic E-field or
dosimetric probe throughout the human phantom (e.g., head, torso or wrist
phantom)~\cite{schmid1996}. Scanning systems are versatile in that the robot can
position the probe anywhere in the phantom and at the required density, avoiding
the need for field reconstruction. Irregular grids are commonly used to locally
improve the resolution in the relevant region of highest SAR. The dosimetric
E-field probe is calibrated in well-controlled induced fields~\cite{meyer2015}.
Robot positioning repeatability is less than 0.05~mm for DASY8 and thus does not
significantly affect the measurement accuracy of the peak spatial-average SAR.

Array systems use a large number of sensors in a fixed grid to electronically
scan the field in the phantom without any moving parts and are therefore much
faster than scanning systems. However, array systems have higher measurement
uncertainty.  Mutual coupling between the sensors limits how close the sensors
can be to each other, which restricts the measurement resolution.  The material
inside the phantom (sensors, transmission lines, and supporting structure),
causes polarization- and distribution-dependent scattering of the induced fields
in the phantom,  making it difficult to fully characterise and remove
scattering-related effects during calibration. Scattering also restricts the
sensors to a single plane conformal to the phantom surface, so that array
systems are dependent on field reconstruction algorithms to estimate the SAR at
locations that cannot be measured. Having no moving parts allows cSAR3D to be
hermetically sealed and use a gel instead of a liquid to prevent separation of
ingredients over time.  The flat cSAR3D phantom studied in this paper has 1024
sensors that measure the three orthogonal field components over a 120~mm x
240~mm measurement area.

\subsection{Requirements for Validation of SAR Measurement Systems}

The requirements for scanning systems and array systems are defined in standards
IEC/IEEE 62209-1528 \cite{iec62209-1528} and IEC 62209-3 \cite{iec62209-3},
respectively.  IEC/IEEE 62209-1528 is based on a standard that was released in
2001 and has been updated several times to account for changes in wireless
technology, measurement system technology, and regulatory requirements. It is
broadly accepted by national regulatory agencies. IEC~62209-3 was released in
2019, but regulators have faced difficulties adopting it. The lack of and need
for formal demonstration of equivalence have been cited as primary concerns
regarding adoption of IEC 62209-3. 

Both standards include validation requirements.  IEC/IEEE 62209-1528
\cite{iec62209-1528} requires validation of each system component separately and
has proven to be very robust. However, such an approach is not possible in IEC
62209-3 \cite{iec62209-3} as array systems are implemented as sealed boxes
without access to the individual components, necessitating a different approach.
Furthermore, array systems inherently have a much larger number of
uncertainty-affecting degrees-of-freedom (and potential sources of failure), due
to the large number of sensors that are independently calibrated for the large
number of power levels, frequencies, and modulations needed to accurately
measure any mobile wireless communication signal. The standards committee was
faced with the issue that a comprehensive set of tests would include millions of
configurations, which cannot be practically implemented. The reduced set of
standardized exposure conditions currently listed in IEC 62209-3 has proven
insufficient, as comparison studies using different array systems show large
deviations in the measured psSAR (as much as 5~dB~\cite{isedcomparison,
watanabe19}). This has prompted the Joint Research Center of the European
Commission to recommend improving the IEC 62209-3 validation \cite{jrc}.

A validation method is needed that satisfies the following
requirements, in order to establish equivalence of the standards and drive
universal acceptance of SAR measurement systems:
\begin{itemize}
\item it is universally applicable to any SAR measurement system (device
    agnostic), 

\item it is able to use knowledge of the measurement system to reduce
    validation effort,

\item it ensures that any successfully validated system performs within the
    reported measurement uncertainty (as prescribed by corresponding standards)
        for any wireless device,

\item it empowers the test lab to confirm the validation independently of the
    system manufacturer,

\item it identifies critical test conditions that maximize the likelihood of
    detecting inadequate measurement device performance,

\item it can be performed with a reasonable effort, to permit re-validation on a
    periodic basis (or whenever the system is relocated, or hardware or software
        components of the system change),

\item it comprehensively covers the space of all relevant exposure conditions
    and configurations, 

\item it is readily extendable as wireless technologies evolve.
\end{itemize}

To simultaneously satisfy the requirements of comprehensive coverage and
reasonable effort, it is necessary to introduce a stochastic component to the
approach. Using a subset of all validation conditions selected by a procedure
involving stochastic elements each time the validation is performed avoids bias
and ensures increasingly comprehensive coverage over time. It also has the added
benefit of preventing system manufacturers from primarily calibrating the system
in view of a priori known validation configurations such that it would pass
the validation but exceed the reported uncertainty.

\subsection{A Three-Step Validation Approach}

This paper presents a three step approach that satisfies all the above
requirements and demonstrates its successful application to both scanning
systems and array systems. At its center is the elaboration of a surrogate model
that estimates the expected measurement error (and the confidence interval
associated with that estimate) of the investigated system for a given exposure
configuration. An important note is that it is a model of the measurement system
error and not of the measurement output itself.  The method consists of three
independent steps, which are each described in detail in corresponding
\textit{Methods} sections:
\begin{enumerate}
\item Model creation: elaboration of the surrogate model using a comprehensive set
    of measurements,

\item Model confirmation: independent confirmation that the surrogate model
    established in Step 1 is valid -- otherwise, the model must be revised,

\item Targeted search for critical cases: search of the configuration space for
    regions with non-negligible likelihood of exceeding the maximum permissible
        error (MPE) using the confirmed model to maximize detection probability,
        and testing the identified configurations to ensure that the measurement
        system performs with the required accuracy.
\end{enumerate}
In the context of SAR measurement system validation, model creation
would typically be performed by the system manufacturer, while
independent model confirmation and the targeted search for critical cases
is typically performed by a test lab or the system user (at least once per year;
the acceptable effort for each party should be within 1\,day).

Gaussian Process (GP) modeling is a classical surrogate modeling approach that
is capable of estimating system responses on a continuous parameter space with
zero-centered error values and of providing associated interpolation
uncertainties based on any given set of known values at specific locations. Such
a model will be referred to as the `GP model' or `surrogate model'. Background
information on geometric GP models can be found in the \textit{Supporting
Information} section. For extended background we refer to, e.g.,
\cite{chiles-delfiner} and \cite{isaaks-srivastava}.

The application of the developed approach to a scanning system and an array
system in the present study demonstrates the feasibility (with acceptable
effort), sensitivity, and generality of the proposed approach. Indeed, the
method is sufficiently general to be of value in a wide variety of applications
beyond SAR systems that require critical, independent, and efficient validation
of system performance.

This method has been proposed for adoption in the next revision of IEC 62209-3
and for a future unified standard that incorporates both and IEC 62209-3 and
current IEC/IEEE 62209-1528.

\subsection{Study Goals}

The principal goal of this study is to identify and demonstrate a practical,
robust, trustworthy, efficient, and comprehensive solution 1) to the specific
problem of SAR system validation and 2) to the more general one of efficiently
and reliably validating systems that are expected to perform within their
uncertainty bounds everywhere in the configuration space.  It is \emph{not} the
goal to develop novel GP modeling theory (other than the extensions needed for
the search algorithms, such as the $\delta_p(l)$ function derived in the
\textit{Supporting Information}), nor to identify the best possible approach to
GP model creation, validation, and exploitation.  The methods applied here were
selected based on accessibility and effectiveness for the task at hand, while
respecting practical SAR measurement constraints such as: a) being non-iterative
(i.e., all measurements are performed in a single session); b) ability to reduce
the continuous validation space to a discrete set of measurable configurations
(e.g., test labs can only have a finite number of validation antenna sources);
and c) respecting the physical parameters of the measurement space (sensor
resolution and accessible sensor region).

\section{Methods}

This section presents the background of the proposed approach, before
introducing the investigated generic application and the real world applications
that build upon the test configuration space of \cite{iec62209-1528,
iec62209-3}.

\subsection{Step 1: Model Creation}

In the first step, a GP model of the deviations from the target values is
constructed from a measured sample set.  The \textit{Supporting Information}
provides background on GP modeling theory, establishes the employed notations,
and defines geometric GP models based on a finite set $S$ of known
configurations in the $n$-dimensional parameter space $X$ in $\Rbb^n$ (whose
measured results are denoted $\bar{S} \subset \Rbb^n \times \Rbb$ with the added
component for the measured values).  Many GP model creation approaches exist and
can be employed, as long as they are capable of conservatively estimating
variances.  For this study, we assume a geometric GP model, based on ordinary
kriging, with the Matheron estimator for the semi-variance, a Gaussian
theoretical variogram model, and the default variogram fitting algorithm from
the SciKit-GStat package \cite{malicke}. These choices were based on the
analysis of the collected data on SAR measurement systems, but alternative
choices are of course possible.

\subsection{Step 2: Model Confirmation}

Given a geometric GP model, a statistical validation procedure is performed in
the second step, after which the model is either rejected or considered
trustworthy. This procedure shall be referred to as the \emph{model
confirmation} procedure; it can be performed as a series of successive tests,
where each success leads to the next test, and the model is considered valid if
and only if the procedure reaches its end. The procedure can be divided into two
main phases: \emph{goodness of fit} and \emph{residuals validation}.
Statistical model validation is a well studied field and valuable information
can be found, e.g., in \cite{roustant2012dicekriging}. The metrics and tests
chosen for this study are commonly employed in statistics.

\paragraph*{Goodness of Fit}

The goal is to assess how well the model's theoretical variogram $\gamma$ fits
the empirical semivariogram $\hat{\gamma} = \{\hat{\gamma}_j\}_{j=1}^k$ built
from the known sample set $\bar{S} = \{(x_i, y(x_i))\} \subset X \times \Rbb$.
For each $j = 1, \ldots, k$ one defines $\gamma_j$ to be the value returned by
$\gamma$ at the lag corresponding to $\hat{\gamma_j}$. As opposed to the often
used mean absolute error (MAE) or root mean square error (RMSE) the
\emph{normalized root mean square error} (NRMSE) is used as its magnitude does
not depend on the sill. 
\begin{align}
    \text{NRMSE} 
    = \frac{\text{RMSE}}{\text{mean}(\hat{\gamma})} 
    = \frac{ \sqrt{\frac{1}{k}\sum_{j} (\gamma_j - \hat{\gamma}_j)^2} }{\frac{1}{k} \sum_{j} \hat{\gamma}_j },
    \label{eq12}
\end{align}
The GP model passes the goodness of fit test when $\text{NRMSE} \leq \alpha$ for a given acceptance value
$\alpha$ (typically in the interval $[0.1, 0.3]$, see \textit{Parameter
Choices}).

\paragraph*{Residuals Validation}

The goal is to evaluate how well a fitted model generalizes to other samples $T
\subset X \subset \Rbb^n$ whose measured set is $\bar{T} \subset X \times \Rbb$.
In practice, such a test sample is obtained either as the test subset of a
train/test cross-validation-partition of a larger known sample, or as a set of
measurements acquired independently from the initial sample that was used to
build the model. The residuals of $T$ against the given model are the
differences between the measured values $y(x)$ and the corresponding predictions
$\hat{y}(x)$. If the residuals appear to behave randomly and follow the expected
distribution, it suggests that the model successfully represents the underlying
data.  On the other hand, if non-random structure is evident in the residuals,
or if their distribution is not the one predicted by the model, it is a sign
that the model poorly fits reality.

The validity of the model is evaluated based on a randomly distributed test sample
$\bar{T} = \{ (x, Y(x)) : x \in T\} \subset \Rbb^n \times \Rbb$ such that:
\begin{itemize}
    \item $T$ is independent from $S$ with $S \cap T = \emptyset$, 

    \item $T$ is locally (stratified) randomly uniformly distributed, in the
        sense that an element of $T$ must be randomly picked within a predefined
        neighborhood (for example within its square if $T$ is generated via
        latin hypercube sampling (LHS) \cite{mckay, park}).
\end{itemize}
This last point differs from $S$, where global evenness (as opposed to local
randomness) is a necessity. The evenness of $T$ is less important, while its
randomness is crucial. The independence between $T$ and $S$ is equally crucial.
In the extremely unlikely case where some configurations $x$ belongs to $S \cap
T$ after generating $T$, they should not be used, or $T$ should be resampled, as
zero residuals are not allowed in the next phase.

For every $x \in X$, the kriging function returns a pair $(\hat{y}(x), \hat{e}(x))$, 
where $Y(x)$ knowing $Y(S)$
follows a normal distribution of mean $\hat{y}(x)$ and variance $\hat{e}(x)^2$.
Because $\hat{y}(x)$ is an unbiased linear combination of the elements of $y(S)$
with $\hat{e}(x)^2$ the variance $\text{Var}(\hat{Y}(x) - Y(x))$, the random
variable 
\begin{align}
\bar{Y}(x)
    = \frac{Y(x) - \hat{y}(x)}{\hat{e}(x)} \sim \Ncal(0, 1)  \label{eq13}
\end{align}
follows the standard normal distribution. Consequently, many well known
statistical tests are applicable. The suggested procedure is as follows:
\begin{enumerate}
    \item Choose a set $T$ such that $T \cap S = \emptyset$ and $T$ contains $50$ elements
        that are evenly distributed across $X$ with a local randomly generated
        identically independently distributed (\textit{iid}) component 
        (i.e. $T$ is obtained through stratified random sampling),

    \item Measure the set $T$ and and build $\bar{T}$ by recording the measured values,

    \item Use the Shapiro-Wilk test (see \cite{shapiro-wilk}) to determine
        the normality of $\bar{Y}(T)$ with acceptance criterion $\alpha = 0.05$,

    \item Make a QQ-plot of $\bar{Y}(T)$ versus $\Ncal(0, 1)$ on the [0.025,
        0.975] inter-quantile range, and compute the slope and location of its 
        linear regression fit,

    \item As the QQ-plot is expected to be linear based on the above step, use the
        QQ-location $\mu$ (the QQ fit value at zero) and the QQ-scale (the QQ fit
        slope) $\sigma$ to determine standard normality with acceptance $|\mu|
        \leq 1$ and $0.5 \leq \sigma \leq 1.5$ (see Sec.~\ref{sec:methodimplementation}).
\end{enumerate}
A bad QQ-location indicates a poorly calibrated model. The QQ-scale is the
standard deviation of the residuals relative to the ideally expected standard
deviation of $\Ncal(0, 1)$ (i.e., $1$). Slope values below the ideal $1$ are an
indication of model conservativeness (overestimation) and thus are more
acceptable than values above $1$ (hence the 1.5 tolerance factor above vs. the
factor of 2 below).  In practice, a non-negligible NRMSE can be the cause of
slope deviation and, therefore, the interpolation error tolerance should
increase with increasing NRMSE.  By studying different failure scenarios (e.g.,
artificially constructed, non-conservative GP models) the formula
\[
    e(x) = \hat{e}(x) \cdot (1 + \text{NRMSE}),
\]
has been heuristically found to be a suitable.

\subsection{Step 3: Targeted Search for Critical Cases}

Given a fully confirmed GP model, the goal of the third step is to search the
$n$-dimensional domain $X$ for regions containing points $x$ that have
non-negligible probabilities for $Y(x)$ to cross a given threshold (see MPE
definition below) and must be added to the test measurements.  These regions are
referred to as \emph{critical regions} and the contained points as
\emph{critical data points}. As $X$ might have a relatively high number of
dimensions, a good GP model might need a relatively large initial sample $S$,
which can lead to substantially slower kriging computations. It is therefore
essential to establish an algorithm that identifies the critical regions of $X$
efficiently.  This requires an algorithm that exploits the available information
-- as contained in the GP model -- to perform its task. The proposed algorithm
relies on the delta function $\delta_p$  (estimate of the shortest distance from
a measured point to points where the conditionally estimated value has a
probability $p$ of exceeding a given level; introduced in the \textit{Supporting
Information}).

The maximum permissible error (MPE) is defined in IEC 62209-3 \cite{iec62209-3}
for each validation configuration $j$ as
\begin{align}
    \mathrm{MPE}_j = 10 \cdot \textrm{log}_{10}(1 + u_{\mathrm{system},j} + u_\mathrm{source}), \label{eq0}
\end{align}
where $u_{\mathrm{system},j}$ is the manufacturer declared expanded uncertainty
(95\,\% confidence interval) of the SAR measurement system for the validation
configuration $j$ (which may not be larger than 30~\% according to
\cite{iec62209-3}), and $u_\mathrm{source}$ is the expanded uncertainty of the
target psSAR value of the validation antenna (which is set to 15~\%
in~\cite{iec62209-3} based on a tolerance evaluation of the antennas). By adding
$u_{\mathrm{system},j}$ and $u_\mathrm{source}$ directly in (\ref{eq0}), IEC
62209-3 treats them as correlated terms, which lowers the validation burden in
comparison to treating them as uncorrelated and combining them as
root-sum-squared~\cite{gum}. Each validation configuration $j$ must satisfy
\begin{align}
\Delta \mathrm{SAR}_j = 
\bigg| 10 \cdot \textrm{log}_{10} \left( \frac{\mathrm{SAR}_{\mathrm{meas},j}} {\mathrm{SAR}_{\mathrm{target},j}} \right) \bigg| \leq \mathrm{MPE}_j, \label{eqmpe}
\end{align}
where $\Delta \mathrm{SAR}_j$ is the error in the measured psSAR and
$\mathrm{SAR}_{\mathrm{meas},j}$ and $\mathrm{SAR}_{\mathrm{target},j}$ are the
measured psSAR and the verified numerical target psSAR value, respectively. If
$\Delta \mathrm{SAR}_j > \mathrm{MPE}_j$ for any configuration $j$, the reported
system uncertainty $u_{\mathrm{system},j}$ is not met. The declared
$u_{\mathrm{system},j}$ can then either be increased (but not larger than
30~\%), or the measurement system is declared to have failed the validation.

Since the measurement system is expected to perform within its uncertainty
bounds everywhere in the configuration space, the inequality in (\ref{eqmpe}) is
enforced for all steps: model creation (where $\Delta\mathrm{SAR}_j \equiv
\bar{S}$), model confirmation  (where $\Delta\mathrm{SAR}_j \equiv \bar{T}$) and
the search for critical cases (here).

\paragraph*{Search Algorithm}

The proposed search algorithm is motivated by heuristic optimization methods,
with an added uncertainty component. Its goal is to identify critical
configurations to be remeasured (search for global extrema), rather than to
estimate the global probability of exceeding the MPE or to improve the surrogate
model. The considerations that went into designing the search algorithm were
that it must:
\begin{itemize}
\item maximize the chance of detecting configurations with a high likelihood of
    exceeding the MPE, while minimising the number of required measurements; 

\item balance the need to spread coverage of the search space, against the need
    to increase the sampling density if either the response surface fluctuates
        strongly, or the predicted measurement error is close to the MPE such
        that even small fluctuations could result in a violation of the MPE; 

\item adhere to the general philosophy of established validation procedures from
    the currently binding standard (point-by-point evaluation according to a
        pass-fail criterion);

\item converge rapidly and be amenable to efficient implementation.
\end{itemize}

The function $\delta_p$ can be used to build an efficient search algorithm for
identifying critical regions in which $Y$ is likely to cross given threshold(s).
Let $S_0 \subset X$ be a finite, evenly distributed sample over $X$.  Let $f$ be
a realization of $Y$ over its domain $X = \Dcal(f)$; it can for example be a
kriging function built from the elements of $S_0$. Let $T_-, T_+ \in \Rbb$
denote two thresholds such that $T_- < T_+$ (for SAR measurement system
validation, $T_-$ = -MPE$_j$ and $T_+$ = +MPE$_j$, as given in
(\ref{eq0})). Given a constant of repulsion $q \in [0, 1]$, 
we have the $m$-iterations search algorithm \ref{alg1}.

\begin{algorithm}
\caption{Search algorithm: Starting from a sample set $S_\ast = S_0$, the
    algorithm iteratively constructs sets $S_\ast \in (S_k)_{k=1}^m$ by moving
    the elements of $S_\ast$ so as to evenly cover
    critical regions with significant probability of exceeding the 
    MPE.}\label{alg1}
\begin{algorithmic}[1]
\Procedure{search}{$S_\ast \subset X$, $f: X \ra \Rbb$,
    $T_-$, $T_+$, $\delta_p: \Rbb_+ \ra \Rbb_+$,
    $q \in [0,1]$, $m \in \Nbb^\ast$}
\State $Y_\ast = f(S_\ast)$ 
\State $T_0 = (T_- + T_+)/2$
\State $d = -1,\ s = 0$
\For{$k = [1, m]$} 
    \State $\alpha = \frac{1}{2k}$ 
    \For{$j = [0, |S_\ast|)$} 
        \State $x_j = S_\ast[j],\ y_j = Y_\ast[j]$
        \If{$y_j > T_0$} 
            \State $s = 1$
            \If{$(y_j > T_+)$} 
                \State $d = \alpha \delta_p(y_j-T_+)$ 
            \Else 
                \State $d = 2\alpha\delta_p(T_+-y_j)$ 
            \EndIf
        \Else 
            \State $s = -1$
            \If{$(y_j < T_-)$} 
                \State $d = \alpha \delta_p(T_--y_j)$
            \Else
                \State $d = 2\alpha \delta_p(y_j-T_-)$ 
            \EndIf
        \EndIf
        \State $S_\ast' = \{x_j\} \cup \{x_j \pm d \overrightarrow{e_i} \in X\}$ 
        \State $Y_\ast' = f(S_\ast')$ 
        \State $S_\ast^{(j)} = S_\ast \backslash \{x_j\}$ 
        \State $D_\ast' = (\min_{x \in S_\ast^{(j)} } \{|x-x'|\})_{x' \in
          S_\ast'}$ 
        \State $h = \mathrm{argmax}_{i} \{
            s (Y_\ast'[i]-T_0) D_\ast'[i]^{\frac{q}{2}} \}$ 
        \State $S_\ast[j] = S_\ast'[h]$, $Y_\ast[j] = Y_\ast'[h]$
    \EndFor
\EndFor
\State \textbf{return} $S_\ast, Y_\ast$
\EndProcedure
\end{algorithmic}
\end{algorithm}

Even though the search algorithm is not strictly speaking an optimization, 
it can be interpreted as a multi-search variant of a heuristic global
optimization algorithm with an uncertainty component that is simultaneously
population and trajectory based:
At the $i$th iteration, the search algorithm produces a new
sample $S_i \subset X$ in such a way that the series $(S_i)$ converges towards a
sample that is evenly distributed over (non connected) critical regions.  An
element of $\{S_i\}$ is denoted $S_\ast$, and $T_\ast$ denotes an element of
$\{T_-, T_+\}$, where $\ast$ acts as a general placeholder.  Programmatically,
$S_\ast$ can as well be seen as a mutable set equal to $S_i$ at the end of
iteration $i$, and implemented as a sequence of elements $S_\ast[j]$ for $j = 0,
\ldots, |S_\ast|-1$. For a small number of iterations $m \geq 1$, algorithm \ref{alg1}
returns a new sample $S_m$ in which points of $S_\ast$ have been moved towards
local extrema of $f$ in such a way that they are evenly spread throughout the
regions that are likely to be close to or beyond the upper and lower thresholds
$T_{\pm}$. The elements of population $S_\ast$ are moved according to two kinds
of forces:
\begin{itemize}
\item A force that pulls $x$ such that $y(x)$ is pulled towards the nearest
    $T_\pm$.  Once $Y(x)$ crosses that threshold, $x$'s velocity decreases and
        the location of the elements of $S_\ast$ start to accumulate in the
        regions surrounding nearby extrema of $Y$. The force is a function of
        $\delta_p(|y(x)-T|)$ for $T$ the closest threshold,

\item A mutually repulsive force with coefficient $q$ that serves
    to spread the elements of $S_\ast$ and quickly decreases with distance. The
        force on any $x_0$ is a function of its distance to the set $S_\ast
        \backslash \{x_0\}$. This prevents points from converging towards the
        same configuration: they should cover regions of interest in a way that
        maximizes their minimal separation. 
\end{itemize}
The choice of $p$ is empirical. It reflects effort-reliability-balance
considerations and ensures that the algorithm detects threshold violations with
a user-defined sensitivity level. The parameter $p$ must be set according to the
degree of smoothness and the rate of variation of $f$ over its domain $X$: if
$p$ is chosen too small, points are overly likely to be classified as
potentially crossing the thresholds $T_{\pm}$, resulting in a too large number
of requested measurements during the targeted search for critical cases. On the
other hand, if $p$ is overly large, the crossing condition might not trigger
rapidly enough to detect sharp peaks in $f$, potentially leading to undetected
regions.

The parameter $q$ is the constant of repulsion.  It is often reasonable to set
$q = p$: with a lower $p$ a finer search is performed and thus a repulsive force
that decreases rapidly with distance is needed. For coarser searches over wider
regions, points should not be too close to each other, and a higher $q$ ensures
that the repulsion acts on longer ranges. Setting $q = 0$ will remove any
repulsion between points.  

A simple, model-agnostic space-filling design (such as LHS) is sufficient to
initiate the algorithm.  The initial sample $S_0$ used for this study is
described in the \textit{Supporting Information}.

The algorithm can be adapted in various ways. In the present application,
knowing that $S_\ast$ starts as a latin hypercube, the choice was made to search
along each dimension of $X$, reducing the chances of multiple points colliding
too quickly.  If $S_\ast$ is instead chosen to be a lattice with high
regularity, one should rather use search trajectories that incorporate a random
element.

Assuming that $f$ also returns its estimation uncertainty for $y(x)$, that is
$f(x) = (\mu_x, \sigma_x)$ where $\sigma_x = 0$ if $f = y$, one can use
algorithm \ref{alg2} to return the two subsets $L_\ast, U_\ast \subset
S_\ast$ of points that have a probability of at least $p$ to cross the
corresponding thresholds.

\begin{algorithm}
\caption{Filter algorithm: Starting from a sample set $S_\ast$ covering the
    critical regions and a GP model $f$, the algorithm partitions $S_\ast =
    L_\ast \cup U_\ast \cup M_\ast$ according to its $p$-likelihood of exceeding
    $T_-$, or $T_+$, or none of them, respectively.}\label{alg2}
\begin{algorithmic}[1]
\Procedure{filter}{$S_\ast \subset X$, $f: X \ra \Rbb \times \Rbb_+$, 
    $T_-$, $T_+$, $p \in [0,1]$}
    \State $L_\ast = \emptyset,\ U_\ast = \emptyset,\ M_\ast = \emptyset$
    \State $a = \Phi^{-1}(p)$ 
    \For{$x \in S_\ast$}
        \State $\mu_x, \sigma_x = f(x)$
        \If{$\mu_x < T_-$ {\bf and} $a \leq \frac{T_- - \mu_x}{\sigma_x}$}
            \State $L_\ast = L_\ast \cup \{ x \}$
        \ElsIf{$\mu_x > T_+$ {\bf and} $a \leq \frac{\mu_x - T_+}{\sigma_x}$}
            \State $U_\ast = U_\ast \cup \{ x \}$
        \Else
            \State $M_\ast = M_\ast \cup \{ x \}$,
        \EndIf
    \EndFor
\State \textbf{return} $L_\ast, U_\ast, M_\ast$
\EndProcedure
\end{algorithmic}
\end{algorithm}

\subsection{Applications}

The developed validation methodology has been applied to (1) an analytic example
where the underlying response is known, for which we assess the ability of the
proposed method to successfully identify critical configurations that exceed
given thresholds, (2) a subspace of the system validation of cSAR3D to gain an
understanding of the performance of the proposed approach, and full system
validation of (3) cSAR3D and (4) DASY8. In applications (3) and (4), care was
taken to ensure that GP model creation was performed independently of the model
confirmation and the targeted search for critical cases by involving two
different measurement laboratories, LAB1 (BNN, India) and LAB2 (IT'IS
Foundation, Switzerland), with different operators and different sets of
equipment. GP model creation was performed by LAB1, and GP model confirmation
and targeted search for critical cases were performed by LAB2.

The set of equipment used in applications (2)--(4) includes a set of validation
antennas (see below), equipment for delivery and monitoring of stable power to
the antennas (signal generators, amplifiers, directional couplers, power
sensors, power meters, cables, and adapters), and hardware for accurate
positioning of the antennas while minimizing influence on the near field
distribution (spacers, holders, and masks). These are further described, with
minimum specifications to ensure high quality results, in the SAR
standards~\cite{iec62209-1528, iec62209-3}.

It is important to emphasize that the goal is \textit{not} to predict psSAR, but
to predict the psSAR measurement error, $\Delta SAR$, and to identify
configurations where that error is likely to exceed the MPE tolerance (i.e., to
identify validation configurations $j$ where (\ref{eqmpe}) is not met). 

\paragraph*{Analytic Sine Wave Example} 

\begin{figure}[!ht]
\centering
\includegraphics[width=1\columnwidth]{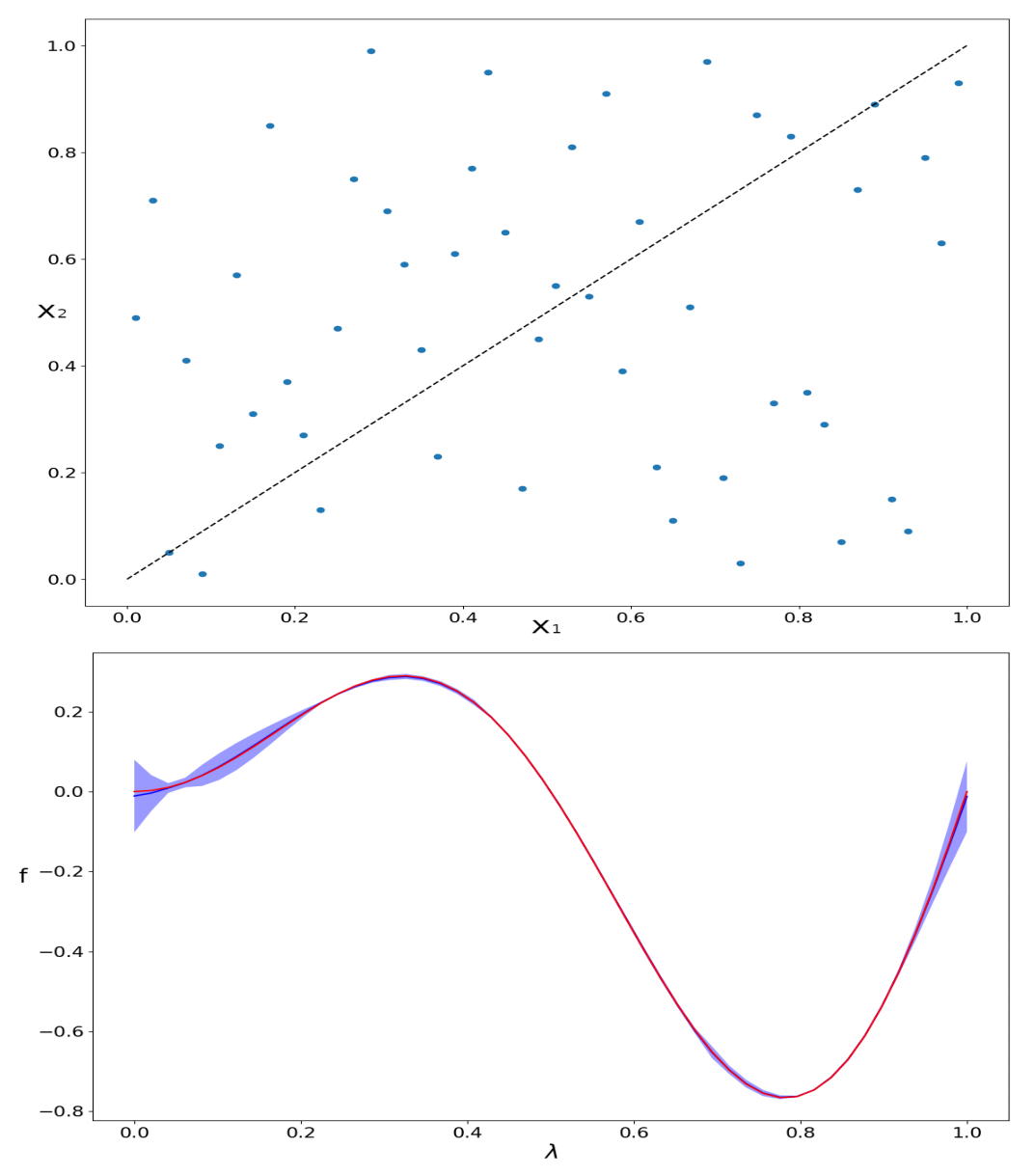}
\caption{Latin square sample $S \in [0,1]^2$ of size 50. The estimated
    interpolation $\hat{Y}(x)$ along the indicated diagonal is shown below:
    The red curve is the exact noiseless function $f$, the blue curve is the
    predicted interpolation $\hat{f}$ along with the associated 99\,\%
    confidence interval.} 
\label{fig3}
\end{figure}

Let $S \subset X = [0, 1]^2$ be the $2$-dimensional LHS sample shown in
Fig.~\ref{fig3}, and let $Y: X \ra \Rbb$ be such that
\[
    Y(x) = f(x) + e, 
\]
where
\begin{align*}
    f(x) &= y \sin(2\pi y),\q
    y = \frac{1}{\sqrt{2}}||x||,\\
    e &\sim \Ncal(0, 0.001^2).
\end{align*}
This process is isotropic and the semivariogram model $\gamma$ is Gaussian with
parameters $r_\gamma = 0.97$, $s_\gamma = 0.22$, $n_\gamma = 0$. Fig.~\ref{fig3} 
shows this model on the segment line $\{\lambda (1, 1) : \lambda \in [0, 1]\}$. 
The confidence interval decreases the closer one gets to a known value, and
increases where the sampling is scarcer and towards the domain's border.

The search algorithms was applied
with varying sensitivities $p$ and iterations $m$, while fixing the threshold
$T_{\pm}$ and reusing $S$ as initial sample set, even though a new (and
frequently larger) sample set would typically be used.

\paragraph*{Dimensions and Range of Parameter Space for cSAR3D and DASY8 Validation}

\begin{figure} \centering
\includegraphics[width=\columnwidth]{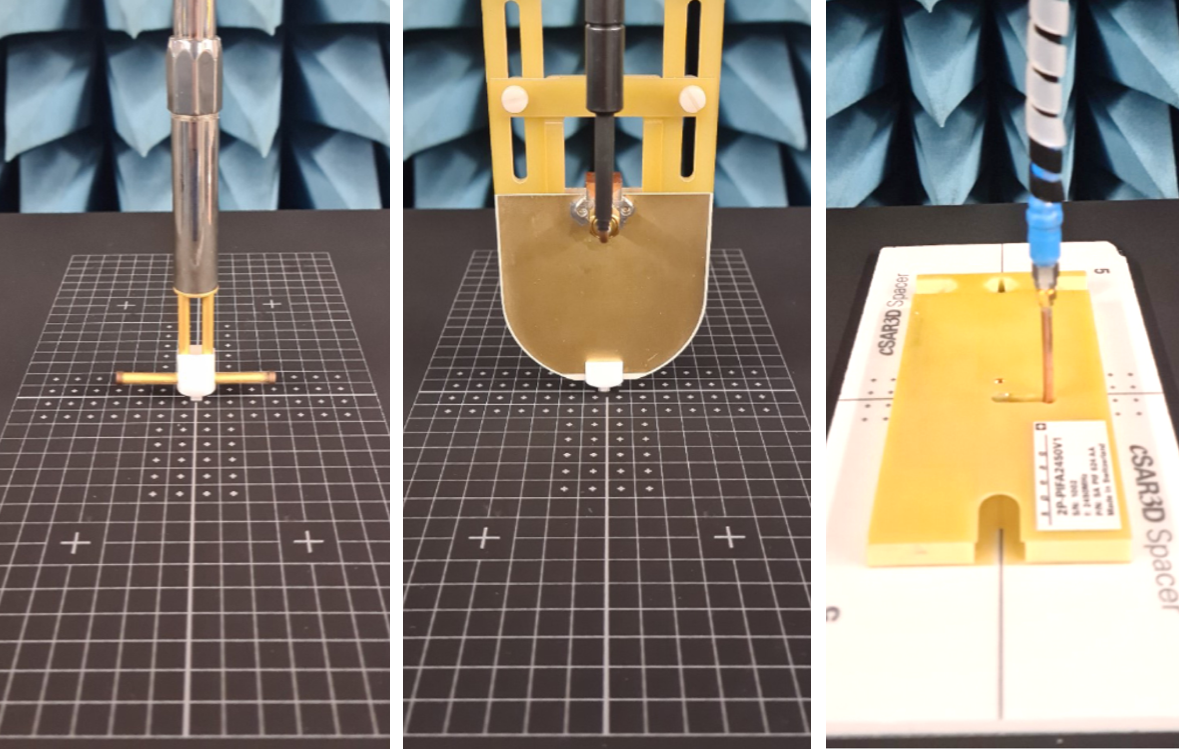}
\caption{Dipole (left), VPIFA (center), and CPIFA (right) validation antennas used in SAR measurement standards \cite{iec62209-3, iec62209-1528}. The dipole and VPIFA test the measurement of parallel and normal polarization with respect to the phantom surface, respectively, while the CPIFA tests the ability to measure multiple peaks.}\label{fig:vpifa}
\end{figure}

To validate a SAR measurement system requires testing of the system using
validation field sources placed in close proximity to the phantom surface. A set
of standardized validation sources is available to the test lab and system
manufacturer lab.  The dimensions of the sampling space must cover all the
relevant variables, and each dimension must cover a range sufficient to validate
the system for all foreseeable exposure conditions. The measurement accuracy of
a SAR measurement system depends on the following SAR parameters (with relevant
validation variables in parentheses):

\begin{description}
\item {\bf Frequency ($f$):} The calibration of the SAR measurement system is
    frequency-dependent, due to the dispersion of the tissue-simulating medium
        dielectric parameters and the frequency response of the system
        components (e.g., amplifiers, filters). Typically, the sensors are
        calibrated at discrete frequencies and the calibration is interpolated
        to cover entire bands. To cover the frequency range of IEC 62209-3 ($f$
        = 300~MHz -- 6~GHz), the validation uses 15 dipole antennas to cover 18
        frequencies. These are 14 dipoles each operating at a single frequency
        labeled D300 (300~MHz), D450, D750, D835, D900, D1450, D1750, D1950,
        D2300, D2450, D2600, D3700, D4200, and D4600, plus a broadband dipole
        labeled D5000 that operates at 4 frequencies (5200, 5500, 5600, and
        5800~MHz).

\item {\bf Polarization ($\theta$):} The system must be able to measure the
    induced E-field for any polarization. A measurement system might do this
        using multiple, e.g., orthogonally-oriented, E-field sensors inside the
        probe that can be combined to isotropically sense arbitrary field
        orientations.  Imperfect isotropy, i.e., angle-of-incidence-dependence,
        affects the measurement uncertainty and needs to be considered in the
        validation. To cover the combined space, different sources and varying
        orientations are defined: 
    \begin{enumerate}
        \item sources with a dominant polarization parallel to the phantom
            surface to test the measurement accuracy of the $x$ and $y$ components. The aforementioned 
            dipole antennas are used for this purpose (see Fig.~\ref{fig:vpifa}). 
        \item sources with a dominant polarization normal to the phantom surface
            to test the $z$-component measurement accuracy. Four VPIFAs
            (vertically-polarized inverted F antennas, see Fig.~\ref{fig:vpifa}) operating at a different frequency each
            and labeled V750 (750~MHz), V835, V1950, and V3700 are designed for this 
            purpose. A PIFA structure~\cite{sanchez2008} is oriented such that the open 
            end is close to the phantom and the short-circuited end is on the opposite end 
            away from the phantom. This provide capacitive coupling at the phantom surface resulting
            in a dominant normal E-field polarization. The evanescent fields decay 
            rapidly, resulting in a sharp SAR distribution and thus providing a good test 
            of the capabilitity of the measurement system. 
    \end{enumerate}
    Antennas should be rotated at all angles $\theta \in [0^\circ, 360^\circ)$  with respect to the 
    phantom surface normal. This can be reduced to $[0^\circ, 180^\circ)$ or even $[0^\circ, 90^\circ)$ 
    without loss of rigor for antennas having SAR patterns with one-fold (VPIFAs) or two-fold (dipoles) 
    reflectional symmetry, respectively. The range $\theta \in [0^\circ, 180^\circ)$ with steps of 
    $\Delta\theta = 15^\circ$ has been used for both antennas for simplicity.
    
\item {\bf SAR pattern ($s$):} The measurement accuracy can depend strongly on
    the spatial distribution of the SAR, due to the field reconstruction
        algorithms, sensor design, sensor spatial resolution, and sensor
        distance to the phantom surface (which affects the capability to
        reliably measure rapidly decaying fields). The measurement system must
        be validated with sources that cover the range of potential SAR
        distributions from wireless devices. This is validated by using
        different antenna types and frequencies, as explained earlier.
        Additionally, dipole antennas are tested at distances from the dipole
        axis to the phantom lossy medium of $s$ = 5, 10, 15, and 25~mm using
        spacers ($s$ = 5~mm is excluded at 300, 450, and 750~MHz due to their
        thicker dipole arms). To test SAR patterns from wireless devices having
        multiple SAR peaks, a dual-peak antenna is used. It is known as the
        CPIFA (centrally-fed inverted F antenna) and operates at 2450 MHz
        (labeled C2450).

\item {\bf SAR level ($P_{in}$):} The measurement accuracy depends on the SAR
    level, due to non linearities of the SAR measurement system that are
    compensated through calibration.  Therefore, the system needs to be
    validated across the system's dynamic range. Input power to each antenna,
    $P_{in}$, is varied over a 20~dB range in 1~dB steps such that SAR$_{1g}$
    varies from 1 -- 100~W/kg for CW (continuous wave) signals, and 0.1 --
    10~W/kg for modulated signals. Reliability should be tested up to an upper
    bound of 100~W/kg for the local SAR, which corresponds to the extremity
    10\,g-averaged SAR limit of 4~W/kg for any SAR distribution.  $P_{in}$
    values depend on the source type, frequency and distance. 

\item {\bf Modulation (PAR, BW):} Sensors are calibrated using different
    modulated signals commonly employed in wireless devices (e.g., 3G, 4G,
    5G, and WLAN signals).  The signal bandwidth, BW (in MHz) is an
    important signal parameter in the frequency domain, while the
    peak-to-average ratio (PAR, in dB) describes signal aspects in the time
    domain. To cover ranges of PAR = 0 -- 12~dB, and BW = 0 -- 100~MHz, a
    set of 22 common modulations has been selected that include two 3G
    signals, ten 4G signals, six 5G signals, and four WLAN signals. An
    unmodulated (CW) carrier has been included to test the upper end of the
    dynamic range, and a pulsed CW signal with a 10~ms period and 10~\% duty
    cycle has been included to test pure pulsed signals.

\item {\bf Location ($x$, $y$):} Array systems must be validated at any $x$ and
    $y$ locations within the measurement boundaries, because the measured SAR
    varies due to a) mechanical tolerances and calibration errors of the
    different sensors in the array, b) measurement variations related to the
    discrete field sampling by the sensors, and c) the influence of field
    scattering across the sensor array that may be different in the center of
    the array compared to the array boundaries. During validation, $x$ and $y$
    are varied over the full measurement area using a 1~mm resolution to include
    locations between the sensors.
\end{description}

These $n = 8$ dimensions ($f, \theta, s, P_{in}$, PAR, BW, $x$, and $y$) are
considered to be a sufficient set for the system validation of general SAR
measurement systems, while six dimensions are sufficient for scanning systems
due to their independence on the device location ($x$, $y$). Knowledge about the
SAR measurement system implementation could be used to further reduce the number
of dimensions and of required validation measurements (improper reduction will
result in failure at the model validation step). For example if the scanning
system implementation is independent of the rotation of the SAR pattern, the
sampling of the rotation angle $\theta$ could be reduced or removed. Or, the
dimensions of frequency $f$ and distance $s$ could be reduced and/or combined
for a high-resolution scanning system that is less sensitive to the SAR pattern.
Still it must be kept in mind that both characteristics of the measurement
system as well as of its reconstruction algorithms must be considered. For this
study, the authors chose to maintain the full dimension space to demonstrate
device dependence on all parameters.  The chosen validation antennas (dipoles,
VPIFAs and CPIFA) are described in detail in \cite{iec62209-3}. 

Note that each of these dimensions is continuous by nature. Reduction to a
finite number of measurable configurations is the result of practical
limitations in a) validation hardware (only specific validation antennas and
spacers available to test labs, for which target values have been determined,
can be used); b) measurement setup (modulated signals must be loaded onto the
signal generator); and c) operator ease-of-use (need for manual positioning and
orientation of the validation antennas).

\paragraph*{cSAR3D $(x, y, \theta)$-subcube}

We examine the case of psSAR measurement error on the $(x, y,\theta)$-subdomain
of a cSAR3D device (fixing the other parameters) to assess the ability of the
proposed approach to detect systematic measurement deviation patterns despite
being agnostic about the measurement system implementation. A single dipole
antenna is used at $f$ = 2450\,MHz. It is fed at a power level of $P_{in}$ =
29.2~dBm with a continuous-wave signal (modulation parameters: PAR = 0~dB, BW =
0~Hz). The dipole is placed at a fixed distance of $s$ = 10~mm on 25 randomly
selected phantom surface locations $L = \{(x_i, y_i)\}_{i \in [1, 25]}$. The
rotation dimension was covered using 12 rotation angles $A = \{j \cdot 15^\circ
: j \in [0, 11] \}$ at each location $(x_i, y_i)$. Therefore, the $(x, y,
\theta)$-subcube consists of 300 points
\[
    S = L \times A = \{ (x_i, y_i, j \cdot 15^\circ) \} \subset X \subset \Rbb^3.
\]

\paragraph*{cSAR3D Validation}

For the full cSAR3D system validation, the entire validation procedure was
performed, this time covering the complete configuration space described above.
A GP model for the full configuration space was constructed  at LAB1 from a set
of cSAR3D flat phantom measurements. 

\textit{Initial sample generation:} An initial sample $S$ of 400 data points
(see \textit{Parameter Choices}) was LHS-generated with a maximization of the
minimal distance between any two points, so as to ensure that $S$ was well
spread within $X$.  For the chosen configurations, the 1g-averaged psSAR
($\text{SAR}_{1g}$) was measured using a cSAR3D device (flat phantom), and the
corresponding deviations $\Delta\text{SAR}_{1g}$ from the published target
values were computed, resulting in the valued sample $\bar{S}$. 

\textit{Outlier detection:} Potential outliers were detected and their
measurement values double checked to eliminate any operator errors.  An
\emph{outlier} is defined to be any value not in the set 
\[
    \left\{\ x \in S\ :\ y(x) \in [q_{1} - rq, q_{3} + rq] \ \right\},
\]
where $q = q_3 - q_1$ for $q_1$, $q_3$ the first and third quartiles of $y(S)$,
and where $r$ is a positive predetermined interquartile range multiplier chosen
equal to $2$ in this case.  With $r=2$, a few valid values might still be
classified as outliers. Outliers are not to be ignored in applying
(\ref{eqmpe}). Nor are they ignored from the linear system of equations used for
interpolation, but they are to be ignored during the construction of the final
isotropic variogram.

\textit{Model creation:} The data space was prescaled based on the standard
deviations of the known values $y(S) = Y(S)$ along each dimension; i.e., for
$Y(S)_i$ the projection of $Y(S)$ on the $1$-dimensional $\Rbb$-subvector space
generated by the $i$-th variable, and for $s_i$ the standard deviation of
$Y(S)_i$, the invertible linear map $\iota_0$ is pre-constructed as the diagonal
matrix
\[
    \Sigma_0 = 
    \begin{pmatrix}
        s_1 &        &   0 \\
            & \ddots &     \\
          0 &        & s_n \\
    \end{pmatrix}.
\]
Working from $\iota_0(X)$ not only normalizes the arbitrary choice of units, but
also reduces the conical uncertainties in the construction of the
$1$-dimensional directional variograms along each dimension (lag pairs are
chosen with an angular tolerance -- set to $45^\circ$ in our implementation --
around the direction-of-interest to improve the statistics).  A Gaussian
theoretical variogram is used for the fitting and, due to the very different
characteristic length of each variable's variation, only the range, sill and
nugget need to be determined along the eight directions of the canonical base
vectors of the surrounding space $\Rbb^n$. The sill was found to be similar
along all directions, so that $\iota$ was defined as the composition
\[
    \Sigma =
    \begin{pmatrix}
        r_1 &        &   0 \\
            & \ddots &     \\
          0 &        & r_n \\
    \end{pmatrix}
    \cdot
    \Sigma_0,
\]
where $r_i$ is the range of the directional variogram along dimension $i$. Now
that $\iota(X)$ is isotropic, an $8$-dimensional semivariogram $\gamma$ on
$\iota(X)$ can be estimated. 

\textit{Model confirmation:} To confirm the model, we first quantify its NRMSE
value to assert that the isotropic semivariogram $\gamma$ fits the empirical
semivariogram $\hat{\gamma}$ sufficiently well  within an acceptance threshold
of 25~\% (see \textit{Parameter Choices}).  Next, the GP model is tested using
measurements taken at LAB2, an independent laboratory with a different operator
and different equipment than those involved in the model creation performed at
LAB1. An appropriate random test sample $\bar{T}$ of 50 points is generated and
the residuals are obtained from the values in $\bar{T}$ using (\ref{eq13}).
Provided $T$ has good randomness, the residuals are to be standard normal
distributed. The Shapiro-Wilk test with a $p$-value well above $0.05$ did assert
normality, and the QQ-plot of the order statistics of the residuals versus the
theoretical standard normal distribution were assessed.

\textit{Targeted search for critical cases}: The confirmed GP model was used to
search the data space for critical regions where psSAR deviations to the target
values ($\Delta\text{SAR}_{1g}$) are likely to exceed the MPE.  A sample of
appropriate size and distribution was LHS-generated and used to initiate the
search Algorithm~\ref{alg1}.  Eight iterations were performed and the
configurations were returned where the probability that
$\Delta\text{SAR}_{1g}$~>~MPE is at least 5\,\%. After the last iteration,
snapping to the closest meaningful neighbour was performed (see
\textit{Methodology and Implementation}). These critical cases were then
measured to check whether or not the values exceed the MPE. 

\textit{Measurement of critical cases}: The identified critical configurations
were measured on a cSAR3D flat phantom to determine whether the measurement
system passed the validation (i.e., $\Delta\text{SAR}_{1g} \leq \text{MPE}$).
This was performed by LAB2 with different equipment and operator than those used
by LAB1 for the model creation measurements.

\paragraph*{DASY8 Validation}

For the full DASY8 system validation, as in the case of cSAR3D, a GP model was
constructed based on system knowledge and an LHS-generated initial set $S$
system knowledge. This use of system knowledge does not compromise the
validation as the model confirmation step, which requires the measurement of a
test set $T$, is carried out independently.  The model creation step was
identical to that of the cSAR3D case, except that a non-zero nugget was
introduced as model parameter, since the nugget was found to be non-negligible
compared to the sill. The model confirmation and the targeted search for
critical cases were conducted for DASY8 in the same way as for cSAR3D.

\subsection{Methodology and Implementation} \label{sec:methodimplementation}

Selected implementation-specific aspects, such as parameter choices, are
discussed here.  

\paragraph*{Parameter Choices}

The proposed procedure involves a number of tuneable parameters.  To establish
the GP model based on a LHS-distributed initial sample, a size of 400 is usually
needed to ensure that $75~\%$ of the 50 bins used in the construction of the
empirical semivariogram contain at least 40 lag values. The Gaussian
semivariogram model was chosen from the smoothness of the underlying process (a
result of the smooth dependence of the measurement physics on variations of the
underlying parameters). It also has the advantage of being less sensitive to
variance changes at the smallest lags, where the LHS generated initial sample
provides few to no values.  A value in the range 10--30~\% is typically chosen
for the NRMSE tolerance in the fit validation of the model confirmation step.
Data analysis has shown that a tolerance below that range is too severe. As
explained in \cite{shapiro-wilk}, the Shapiro-Wilk test is best applied to
samples that have 20-50 elements, while at least 50 points are recommended for
the QQ-plot to be meaningful. The usual tolerance of 5\,\% is applied; this
could easily be increased for more severity, however, normality itself is less
important than the scale and location factors of the QQ-test.  The QQ-test
location and scale tolerance are set at $|\mu| < 1$ and $0.5 \leq \sigma \leq
1.5$ after normalization to the standard deviations (\ref{eq13}). In order to
accept more conservative models, the tolerance for $\sigma$ is more permissive
below than above the ideal $1$.  The constant of repulsion $q$ in the search
algorithm should typically be chosen in the range of 0.05--0.2, depending on the
expected global smoothness of $\Delta\text{SAR}$ over the configuration space. A
default value of 0.1 has been found to work for the types of measurement systems
studied, The number of iterations $m$ in the search algorithm is set to $8$, as
this proved sufficient for the configuration samples to converge to  the
critical regions.  Finally, the initial search population size is in the range
50--10000.  The search algorithm can efficiently handle such a potentially large
number of trajectories.

\paragraph*{Latin Hypercube Sampling Implementation}

In order to efficiently sample $S$ for the initial GP model creation, the
choice made was to use \emph{Latin Hypercube Sampling (LHS)} (see \cite{mckay}
and \cite{park}). For a unified LHS procedure to generate suitable sample sets
$S$ of size $k$ for both the model creation and the
model confirmation step, the following conditions need to be satisfied:
\begin{itemize}
    \item The cube $I_n = [0, 1]^n$ is partitioned into the canonical grid of
        $k^n$ equally sized $n$-dimensional sub-cubes (`cases'). The elements of
        $S$ are placed in their own `case', such that each row along each of the
        $n$ dimensions contains exactly one element of $S$, and that the minimal
        euclidean distance between two occupied squares is maximized,

    \item each element of $S_0$ is placed within its `case' according to a
        uniform probability distribution.
\end{itemize}
While the first condition guarantees a good initial sample, the second
condition is essential for it to constitute an appropriate test sample.
A custom developed implementation of LHS
based on pyDOE \cite{baudin2013pydoe} was used for this purpose.

\paragraph*{Data Snapping}

Not all configurations in domain $X$ are valid (i.e. physically measurable).
The general approach is to treat $X$ as a continuous connected subset of
$\Rbb^n$, and to derive meaningful values \textit{a posteriori} (after the last
iteration) through snapping to their closest meaningful neighbour.  In this way
the search algorithm does all its works on $X$, after which a final kriging
round is performed to the snapped configurations. In rare instances, if an
element is too far from a meaningful location, it is removed from the resulting
critial set. This approach avoids having to treat categorical variables,
discrete variables and continuous variables differently in the validation
procedure, and greatly simplifies all statistical operations without
significantly affecting the results. 

As for initial samples generation, the source selection is based on the 
(frequency, distance)-pair. If a frequency is valid for different source types, 
distance is used as a secondary criterion. When no meaningful source exists, 
the sample point is ignored.

\paragraph*{Variogram Modeling}

All semivariograms were generated using the scikit-gstat 1.0 package from the
pypi repository; all details on the methods used are provided in \cite{malicke}.
The Matheron semivariance estimator given in (\ref{eq3}) is used for the
empirical variogram construction. The Gaussian semivariogram model defined in
(\ref{eq5}) was found to be the most effective model. The convex nature of the
model at short ranges reduces the fitting uncertainty associated with the
typically sparse sampling data at the shortest lags.  Most importantly the
monotonicity of the semivariogram model ensures that the delta function is
applicable. The domain is binned such that $75~\%$ of its diameter is
partitioned into 25 equally-sized bins for the $1$-dimensional directional
variograms and into 50 bins for the final isotropic variogram. While zero nugget
models worked for cSAR3D, it became obvious that nugget-inclusion was necessary
in the DASY8 case. 

\section{Results}

\subsection{Analytic Sine Wave}

\begin{figure}[!htb]
\centering
\includegraphics[width=1\columnwidth]{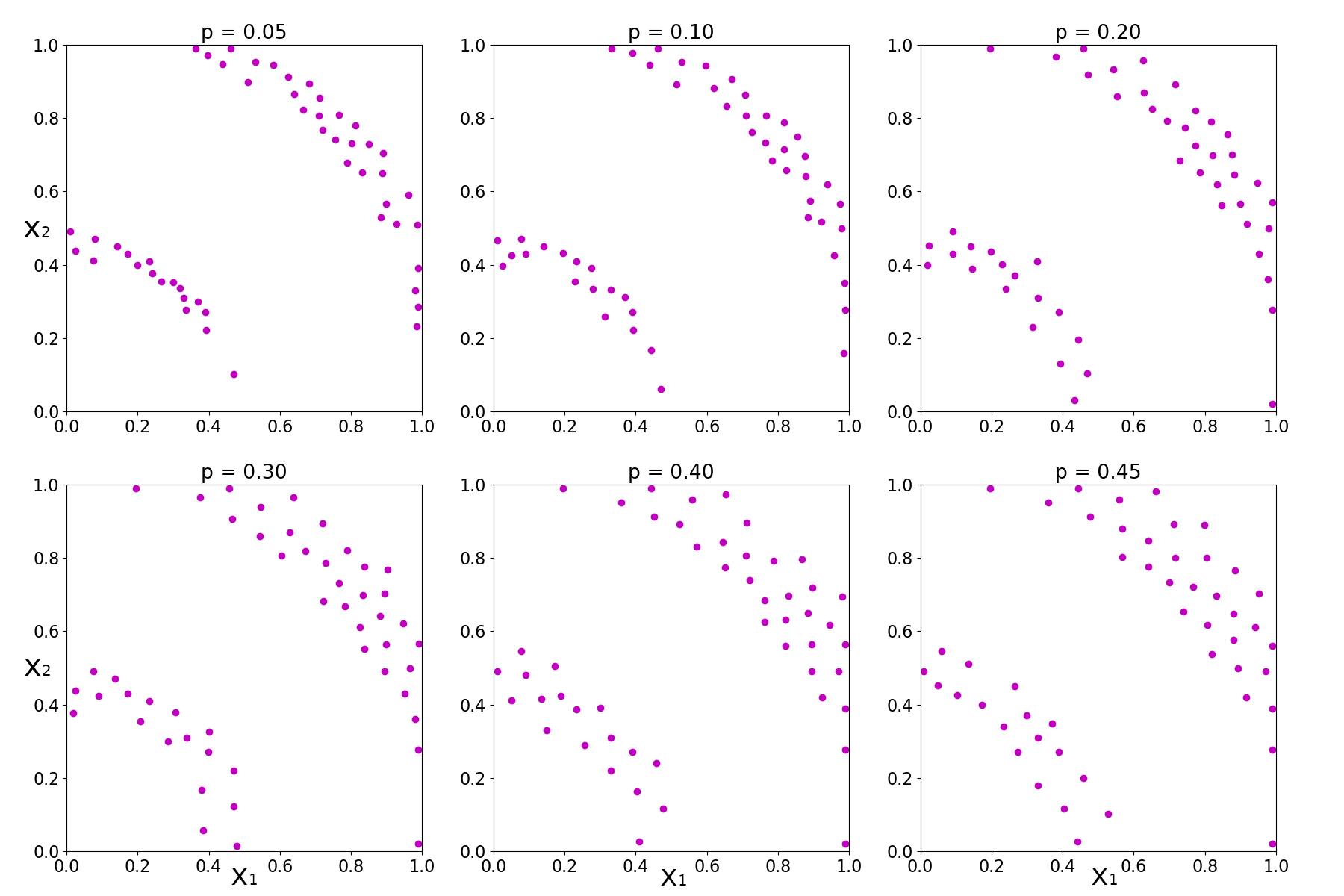}
\caption{Results obtained using Algorithm~\ref{alg1} for the
    targeted search for critical cases. Identified candidate configurations are shown for various values of the
    sensitivity $p$ (always with $m=8$ iterations). A higher $p$ reflects the
    expectation of a smoother response surface, which permits to increase the
    spacing between sampling points and to search the configuration space more
    sparsely.} \label{fig6}
\end{figure}

\begin{figure}[!htb]
\centering
\includegraphics[width=1\columnwidth]{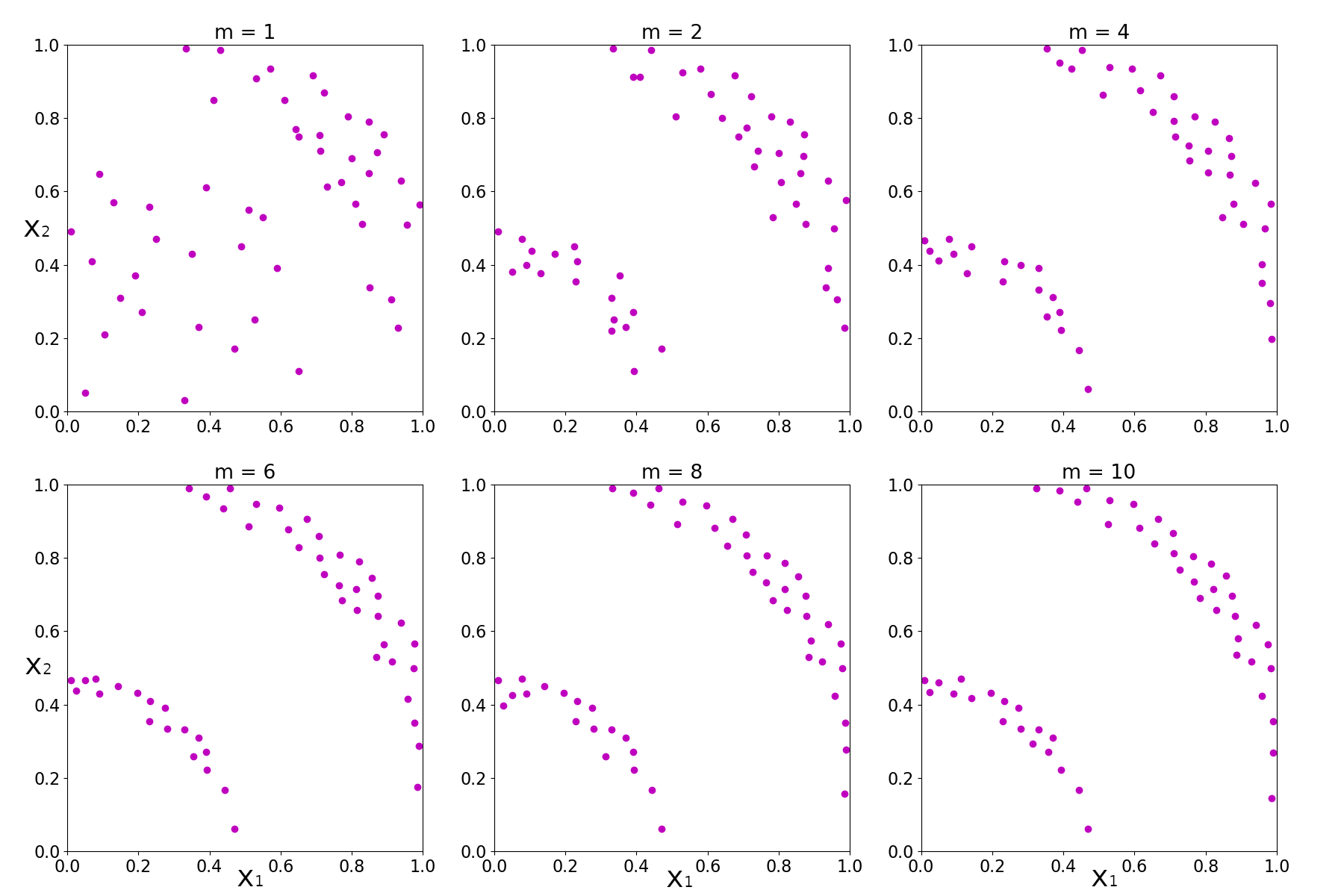}
\caption{Impact of the iteration number $m$ on the results returned by
    Algorithm~\ref{alg1}. The sensitivity parameter is set to
    $p=0.1$ throughout with an increasing number of iterations $m$. Quick
    convergence up to $m=8$, then stabilization beyond $m=8$ of the identified failure candidate is observed.} \label{fig7}
\end{figure}

The application of the targeted search for critical cases (using $T_+$ = 0.20
and $T_-$ = -0.75) for various sensitivities $p$, and iterations $m$, moves the
elements of $S$ as illustrated in Fig.~\ref{fig6} and Fig.~\ref{fig7}
respectively.  The impact of the $p$ value is illustrated in Fig.~\ref{fig6}.
The fact that $\delta_p$ incorporates the relevant semivariogram characteristics
allows the search algorithm to remain efficient with a minimal number of
iterations (provided the semivariogram model is of sufficient quality).
Fig.~\ref{fig7} illustrates how after only two iterations the points have
already converged to cover the regions-of-interest. These figures show the
remarkable regularity in how points are set apart from each other. 

\begin{figure}[!htb]
\centering
\includegraphics[width=1\columnwidth]{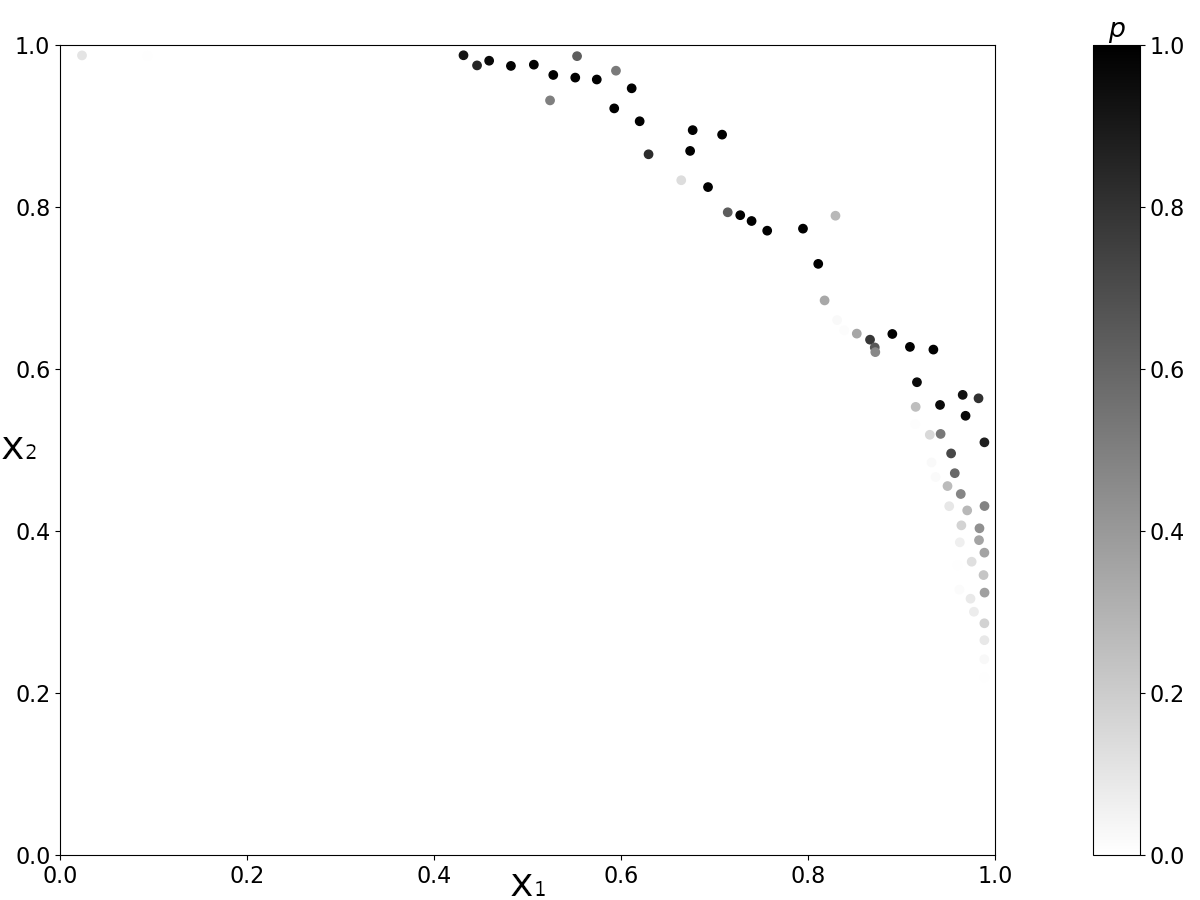}
\caption{Configurations for remeasurement, identified by the complete search procedure
    (algorithms~\ref{alg1} and \ref{alg2}). The probability of the
    samples to fall below the lower threshold of $T_{-}=-0.75$ is indicated.} \label{fig8}
\end{figure}

Each of these points come with their own probability to cross the thresholds.
Fig.~\ref{fig8} shows these probabilities for the lower threshold
of $T_- = -0.75$, which is identical to the infimum of $f$ on $X$. 

\subsection{cSAR3D $(x, y, \theta)$-subcube}

\begin{figure}[!htb]
\centering
\includegraphics[width=1\columnwidth]{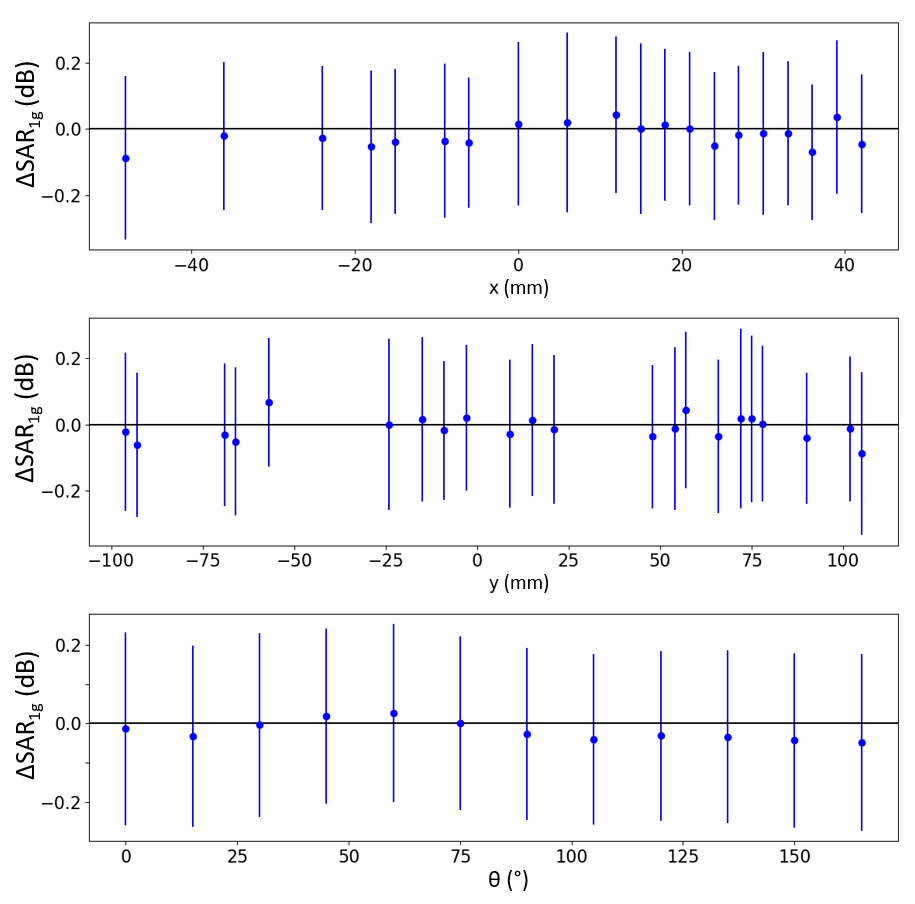}
\caption{The set $\bar{S}$ of the measured cSAR3D deviations
    from the target values, $\boldsymbol{\Delta}\text{SAR}_{1g}$, projected along the different dimensions of the $(x,
    y, \theta)$-subcube configuration space. The bars denote the standard
    deviations of all elements of $\bar{S}$ with the same projected parameter
    value.} \label{fig9}
\end{figure}

The results from applying the developed procedure to the $(x, y,
\theta)$-subdomain of a cSAR3D device provide valuable insights into how regions
can be identified that -- as a result of the measurement system design (in this
case, the arrangement of sensors in the array) -- have an increased likelihood
of exceeding the accuracy limits.  The set of measured configurations $\bar{S} =
\{ (x_k, y_k, \theta_k, \Delta\text{SAR}_k) \}$ can be represented by projecting
all measurement errors  $\Delta\text{SAR}_k$ onto each dimension: their mean and
standard deviation are shown in Fig.~\ref{fig9}. 

\begin{figure}[!htb]
\centering
\includegraphics[width=1\columnwidth]{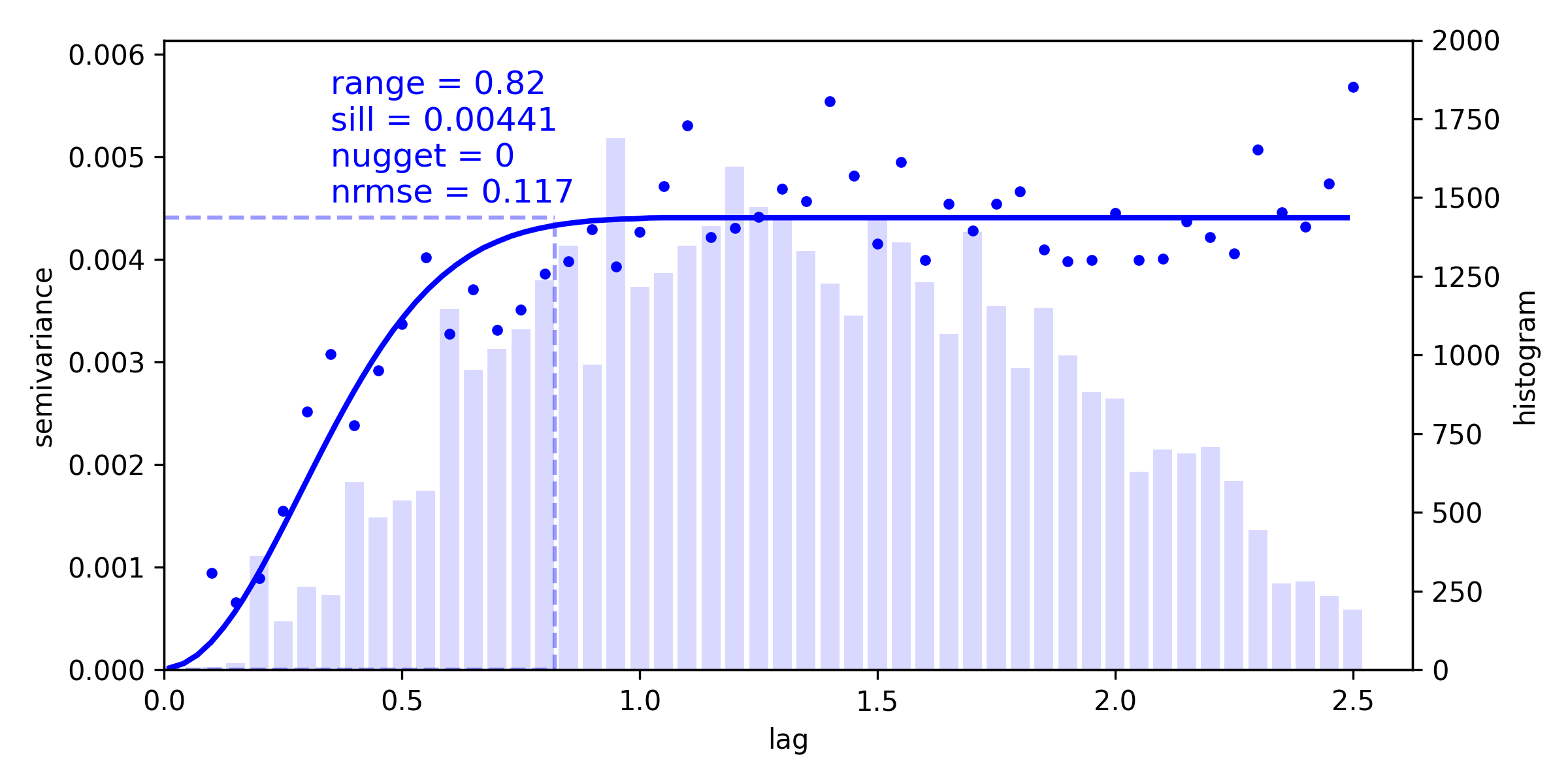}
\caption{Empirical (dots) and fitted (line) isotropic
    semivariograms for the cSAR3D system in the $(x, y, \theta)$-subcube of the
    configuration space. It is shown along with the histogram (bars) of the sample
    lags in the underlying bins.} \label{fig10a}
\end{figure}

Note that the locations are not as optimally distributed (i.e., locally random,
but globally homogeneous) as if they would have been LHS-generated. This is
compensated by the high number of elements in $S$ and the high regularity of all
standard deviations across the board. The data shows a smooth geometric
anisotropy with low noise. By definition of geometric GP models as given in
(\ref{eq7}), one can compute an isomorphism $\iota$ via which the underlying
data space $X$ can be rescaled into an isotropic space $\iota(X)$. An isotropic
semivariogram $\gamma$ built on $\iota(X)$ is shown in Fig.~\ref{fig10a}.

\begin{figure}[!htb]
\centering
\includegraphics[width=1\columnwidth]{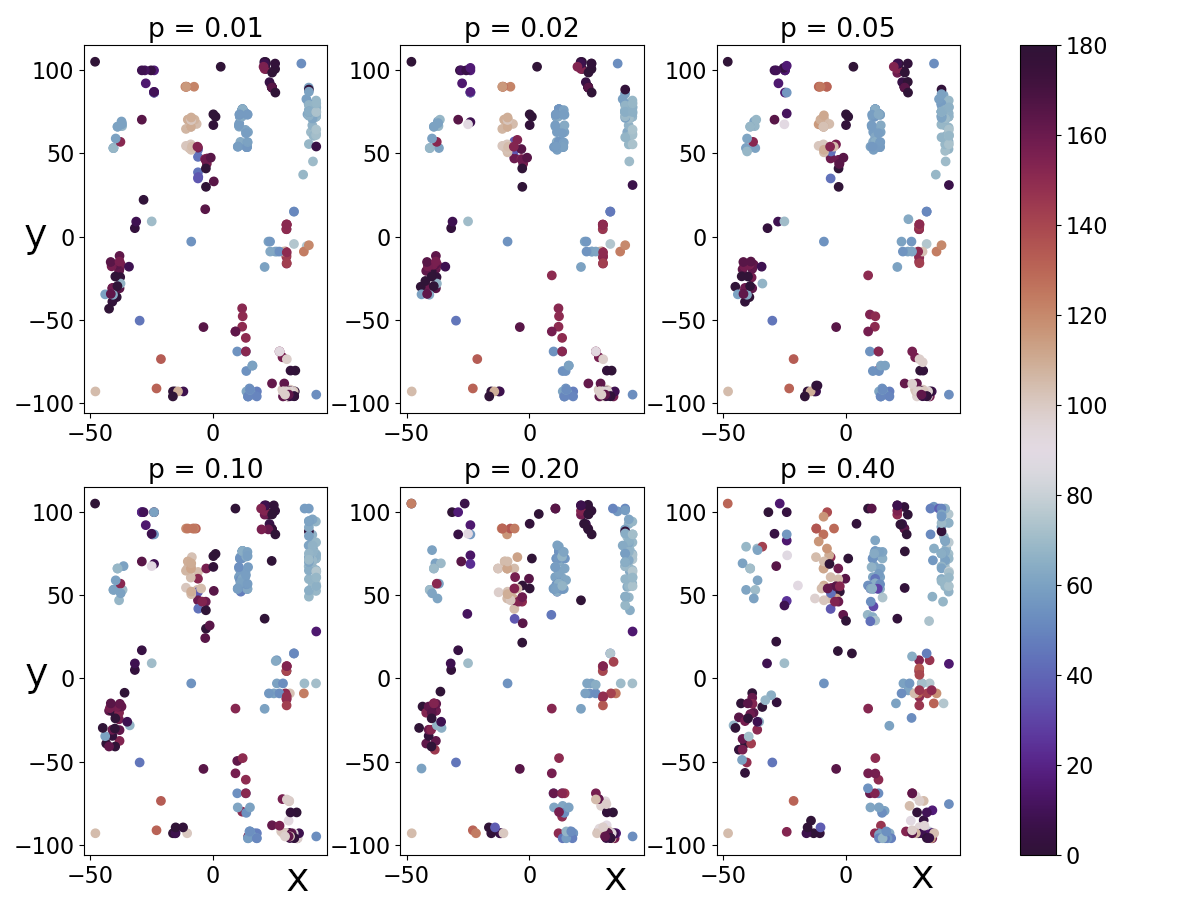}
\caption{Results obtained for the reduced cSAR3D dataset when
    applying Algorithm~\ref{alg1} in the targeted search for critical
    cases. The outcomes for various values of the sensitivity
    parameter $p$ are shown on the $(x, y)$-surface, while encoding $\theta$ in
    color.} \label{fig11}
\end{figure}

\begin{figure}[!htb]
\centering
\includegraphics[width=1\columnwidth]{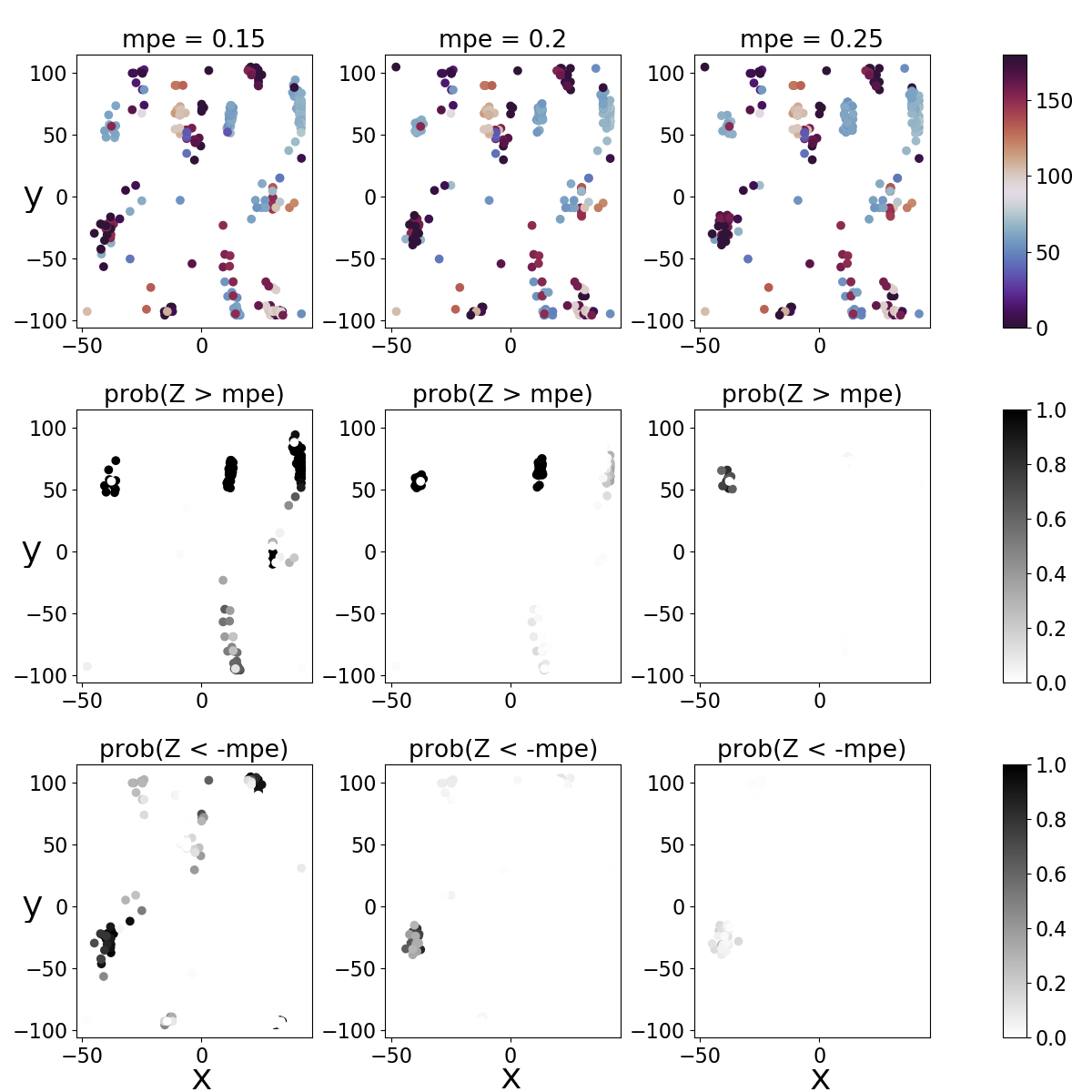}
\caption{Results obtained for the reduced cSAR3D dataset when
    applying both algorithms of the targeted search for critical cases.
    Shown are the outcomes of Algorithm~\ref{alg1} (top; see as well
    Fig.~\ref{fig11}), and of Algorithm~\ref{alg2} (middle and bottom) for 
    an error threshold (0.15, 0.2, or 0.25\,dB) which is well
    below the typical MPE of 1.5~dB or more.} \label{fig12}
\end{figure}

The NRMSE of less than $12~\%$ -- well below the 25\,\% acceptance limit --
indicates that the semivariogram is well suited for probing $X$. Applying the
search algorithm with a very low MPE threshold of 0.3\,dB (well below that
allowed by the standards; see (\ref{eq0})) using different sensitivity values
$p$ results (after only $8$ iterations) in the critical samples shown in
Fig.~\ref{fig11}. Color clustering is apparent in Fig.~\ref{fig11}, which
corresponds to the local extrema along the $\theta$ dimension from
Fig.~\ref{fig9}, namely $15^\circ, 65^\circ, 105^\circ, 120^\circ,$ and
$165^\circ$.  Fig.~\ref{fig12} then applies Algorithm~\ref{alg2} and returns the
probability of exceeding the MPE value. Clear clustering around the global
maximum at $\theta = 65^\circ$ and the global minimum at $\theta = 165^\circ$ is
evident, in addition to spatial clustering.

\subsection{cSAR3D Validation}

\begin{figure}[!htb]
\centering
\includegraphics[width=1\columnwidth]{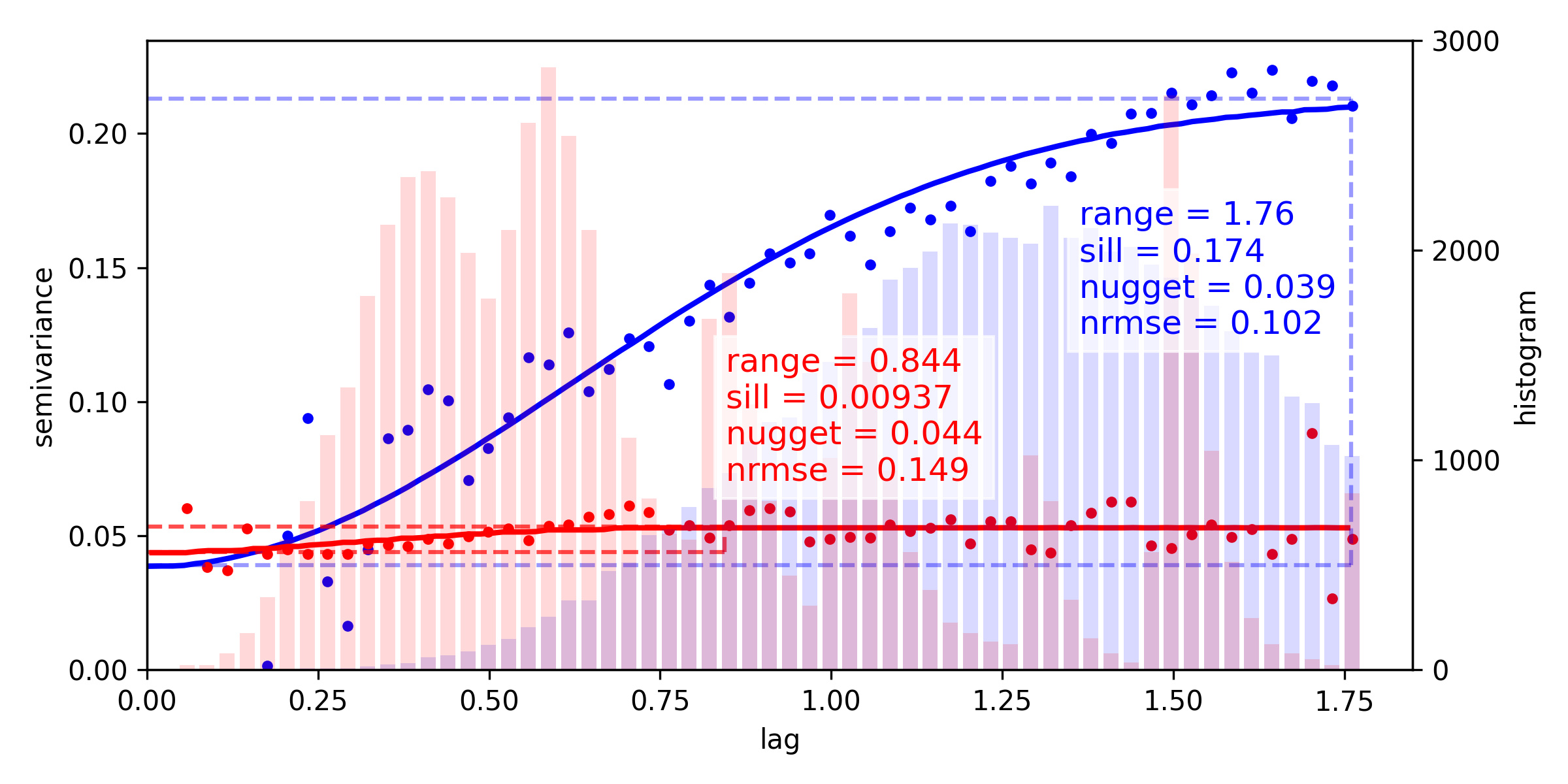}
\caption{Isotropic semivariogram of the cSAR3D (blue) and DASY8 (red) 
    GP models, showing the empirical (dots) and fitted (lines) semivariances along with the histograms (bars) of the binned sample lags. The fit quality for the cSAR3D and DASY8 models returned an NRMSE 
    of about 10\,\% and 15~\%, respectively.}
    \label{fig13a}
\end{figure}

The isotropic semivariogram $\gamma$ fits the empirical semivariogram
$\hat{\gamma}$ with an NRMSE value of about 10~\%, well below the acceptance
threshold of 25~\%.  Typically, a good model has an NRMSE in the range of 8~\%
to 15~\%, which is the case here as shown in Fig.~\ref{fig13a}.

\begin{figure}[!htb]
\centering
\includegraphics[width=0.8\columnwidth]{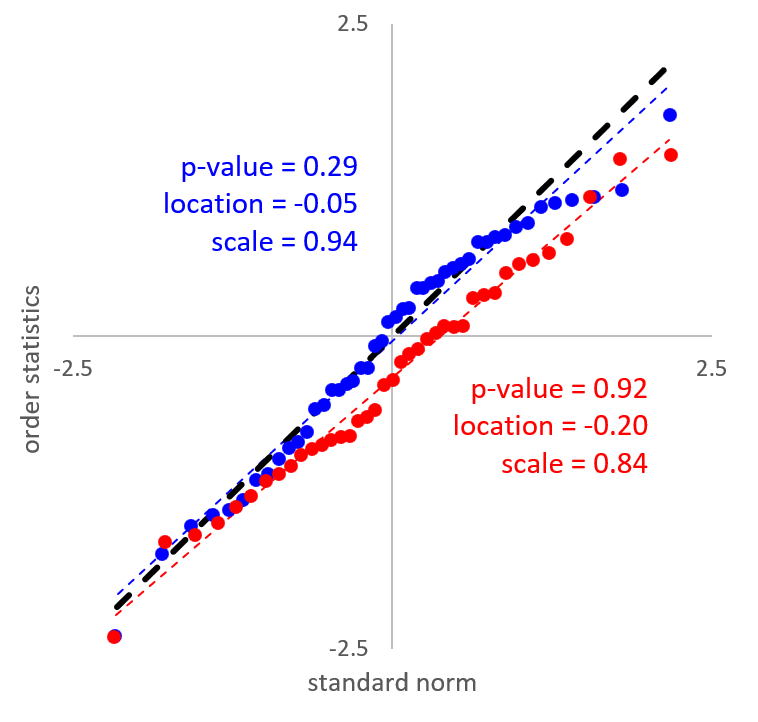}
\caption{QQ-plot of the model residuals (dots) and linear fit (lines) for cSAR3D (blue) and DASY8 (red) shown in comparison to the ideal fit (black line). 
    The Shapiro-Wilk test p-value, as well as the location and scale of
    the linear regression are within their acceptance ranges, confirming the GP
    model.} \label{fig14a}
\end{figure}

The Shapiro-Wilk test confirms that the residuals are indeed normally
distributed, with a $p$-value of $0.29 > 0.05$. As shown in
Fig.~\ref{fig14a}, the linear regression of the QQ-plot of the residuals
order statistics versus the theoretical standard normal distribution has its
location and scale well within $[-1, 1]$ and $[0.5, 1.5]$, respectively. 

The confirmed GP model was then used to search the data space.  The critical
configurations returned after 8 iterations of the search algorithm with an
attributed probability of at least 5\,\% of exceeding the MPE value are listed
in Table~\ref{tab:csar3}. These 44 configurations are sorted by decreasing
probability of exceeding the MPE (from 18.8~\% to 5.1~\%).  A cluster of cases
using the D5000 dipole at a distance $s$~=~25\,mm with input power levels around
$P_{in}$~=~10\,dBm is evident, as well as two cases using the V750 antenna. This
is unsurprising, as the D5000 and V750 antennas have the sharpest SAR
distributions among the sets of dipole antenna and VPIFA, respectively. A
sharper SAR distribution results in a larger measurement variability for an
array system with a fixed sensor resolution.

\begin{table}[ht]\small \centering
\begin{adjustbox}{width=1\columnwidth, center}
\begin{tabular}{|lrcrcccrrrrr|}\hline
&  &  &  &  &  &  & &  &  $\Delta\text{SAR}$ &  \multicolumn{1}{c}{Model}  &  \multicolumn{1}{c|}{Fail} \\
\multicolumn{1}{|c}{Ant.} &  \multicolumn{1}{c}{$f$} &  \multicolumn{1}{c}{$P_{in}$} &  PAR &  \multicolumn{1}{c}{BW} &  \multicolumn{1}{c}{$s$} &  \multicolumn{1}{c}{$\theta$} & \multicolumn{1}{c}{$x$} &  \multicolumn{1}{c}{$y$} &  \multicolumn{1}{c}{1g} &  \multicolumn{1}{c}{error}  &  \multicolumn{1}{c|}{prob.} \\
\multicolumn{1}{|c}{name} &  (MHz) &  (dBm) & (dB) &  (MHz) &  (mm) & ($^\circ$) &  (mm) &  (mm) &  \multicolumn{1}{c}{(dB)} & (dBm) &  \multicolumn{1}{c|}{(\%)} \\\hline
V750 &        750 &    9 &   3.98 &        5 &         2 &     90 & -33 &   32 & -1.249 &  0.282 &  18.8~\% \\
D5000 &       5600 &   11 &   8.91 &       40 &        25 &      0 & -46 &  -60 &  1.102 &  0.362 &  13.6~\% \\
V750 &        750 &    9 &   5.67 &       20 &         2 &     75 & -19 &   22 & -1.161 &  0.284 &  11.6~\% \\
D5000 &       5500 &   10 &   8.91 &       40 &        25 &     15 & -46 &  -38 &  1.076 &  0.352 &  11.4~\% \\
D5000 &       5500 &   13 &   8.91 &       40 &        25 &     15 & -36 &  -68 &  1.102 &  0.329 &  11.3~\% \\
D5000 &       5200 &   10 &   8.91 &       40 &        25 &      0 & -26 &  -18 &  1.104 &  0.318 &  10.6~\% \\
D5000 &       5500 &   10 &   8.43 &       25 &        25 &     15 & -30 &  -19 &  1.108 &  0.313 &  10.5~\% \\
D5000 &       5500 &   10 &   8.91 &       40 &        25 &     30 & -35 &   -5 &  1.092 &  0.326 &  10.5~\% \\
D5000 &       5500 &   18 &   7.93 &        0.4 &        25 &     15 & -43 &   59 &  1.056 &  0.352 &  10.4~\% \\
D5000 &       5600 &   13 &   8.43 &       25 &        25 &     15 & -22 &  -53 &  1.106 &  0.300 &  9.4~\% \\
D5000 &       5800 &   16 &   9.38 &       20 &        25 &     45 & -14 &  -85 &  1.084 &  0.313 &  9.2~\% \\
D5000 &       5500 &   11 &   8.91 &       40 &        25 &     15 & -12 &  -80 &  1.077 &  0.316 &  9.0~\% \\
D5000 &       5800 &   15 &   8.43 &       25 &        25 &     30 &  -6 &  -89 &  1.093 &  0.302 &  8.9~\% \\
D5000 &       5500 &   11 &   9.38 &       20 &        25 &     15 & -31 &   53 &  1.063 &  0.323 &  8.8~\% \\
D5000 &       5600 &   15 &   6.59 &       10 &        25 &      0 & -34 &  -44 &  1.094 &  0.299 &  8.7~\% \\
D5000 &       5800 &    8 &  10.28 &      100 &        25 &     15 &  -7 &  -28 &  0.969 &  0.388 &  8.6~\% \\
D5000 &       5800 &    8 &   8.90 &       80 &        25 &     45 & -15 &  -40 &  1.022 &  0.335 &  7.7~\% \\
D5000 &       5500 &   10 &  10.28 &      100 &        25 &     30 & -13 &    3 &  0.954 &  0.379 &  7.5~\% \\
D5000 &       5600 &   10 &  10.28 &      100 &        25 &     60 &  -1 &   14 &  0.950 &  0.380 &  7.4~\% \\
D5000 &       5800 &   23 &   8.43 &       25 &        25 &     15 & -48 &  -32 &  0.994 &  0.345 &  7.1~\% \\
D5000 &       5800 &    8 &   8.90 &       80 &        25 &     30 & -26 &   72 &  0.904 &  0.406 &  7.1~\% \\
D5000 &       5500 &   10 &   8.91 &       40 &        25 &     15 & -43 &   68 &  0.973 &  0.357 &  7.0~\% \\
D5000 &       5800 &   11 &  10.28 &      100 &        25 &     30 &  -7 &   78 &  0.860 &  0.431 &  6.9~\% \\
D5000 &       5500 &   10 &  10.28 &      100 &        25 &     60 &   2 &   -7 &  0.950 &  0.368 &  6.7~\% \\
D5000 &       5800 &    8 &   8.90 &       80 &        25 &      0 &  15 &  -59 &  0.964 &  0.358 &  6.7~\% \\
D5000 &       5500 &   11 &   8.90 &       80 &        25 &     15 & -38 &   -1 &  0.967 &  0.355 &  6.6~\% \\
D5000 &       5200 &   10 &  10.28 &      100 &        25 &     60 & -18 &  -16 &  0.964 &  0.355 &  6.6~\% \\
D5000 &       5500 &   10 &  10.28 &      100 &        25 &     30 &   2 &  -23 &  0.942 &  0.364 &  6.3~\% \\
D5000 &       5800 &   17 &   6.59 &       10 &        25 &     45 & -44 &   57 &  1.001 &  0.325 &  6.3~\% \\
D5000 &       5600 &   10 &   8.90 &       80 &        25 &     30 & -20 &  -53 &  1.016 &  0.315 &  6.2~\% \\
D5000 &       5600 &   10 &  10.28 &      100 &        25 &     15 &   3 &  -36 &  0.936 &  0.367 &  6.2~\% \\
D5000 &       5600 &   10 &  10.28 &      100 &        25 &     30 &  14 &   -5 &  0.903 &  0.386 &  6.1~\% \\
D5000 &       5500 &   10 &  10.28 &      100 &        25 &     30 &   4 &  -46 &  0.952 &  0.355 &  6.1~\% \\
D5000 &       5500 &   10 &   8.90 &       80 &        25 &     30 &  -4 &    2 &  0.985 &  0.333 &  6.1~\% \\
D5000 &       5600 &   10 &   8.91 &       40 &        25 &     30 & -26 &   39 &  1.005 &  0.317 &  5.9~\% \\
D5000 &       5500 &   10 &   8.90 &       80 &        25 &     30 & -11 &  -13 &  0.996 &  0.322 &  5.9~\% \\
D5000 &       5500 &   10 &  10.28 &      100 &        25 &     75 &   4 &  -63 &  0.928 &  0.362 &  5.7~\% \\
D5000 &       5600 &   10 &  10.28 &      100 &        25 &     45 &  -8 &  -73 &  0.941 &  0.353 &  5.6~\% \\
D5000 &       5500 &   17 &   5.67 &       20 &        25 &     45 &  10 & -105 &  1.043 &  0.287 &  5.6~\% \\
D5000 &       5200 &   10 &   8.90 &       80 &        25 &     15 & -14 &   20 &  0.945 &  0.348 &  5.5~\% \\
D5000 &       5600 &   10 &   8.91 &       40 &        25 &     30 &  -9 &  -23 &  1.024 &  0.297 &  5.5~\% \\
D5000 &       5200 &   10 &   8.90 &       80 &        25 &      0 & -14 &  -21 &  0.961 &  0.334 &  5.3~\% \\
D5000 &       5200 &   10 &   8.90 &       80 &        25 &     30 &  -7 &   19 &  0.946 &  0.340 &  5.2~\% \\
D5000 &       5600 &   10 &   8.90 &       80 &        25 &     45 &  12 &  -34 &  0.975 &  0.321 &  5.1~\% \\\hline
\end{tabular}
\end{adjustbox}
\caption{\label{tab:csar3} 44 identified critical configurations for the cSAR3D system where there is at least a $5~\%$ probability 
    (column labeled \textit{Fail prob.}) that the inequality in (\ref{eqmpe}) is not met. Each configuration is identified by its 8-dimensional coordinate in the validation parameter space ($f, P_{in}$, PAR, BW, $s, \boldsymbol{\theta}, x, y$), while \textit{Ant. name} is the validation antenna operating at the listed $f$ and $s$ combination).
    The three right-most columns are estimated from the GP model: the GP model error ($\boldsymbol{\Delta}\text{SAR}_{1g}$); the standard deviation of the GP model error (\textit{Model error}), and the failure probability (\textit{Fail prob.}).
    Since the MPE of (\ref{eq0}) and (\ref{eqmpe}) is only known after the
    measurement is performed, it was conservatively estimated
    from the MPE values reported during model creation.}
\end{table}

\begin{figure}[!htb]
\centering
\includegraphics[width=1\columnwidth]{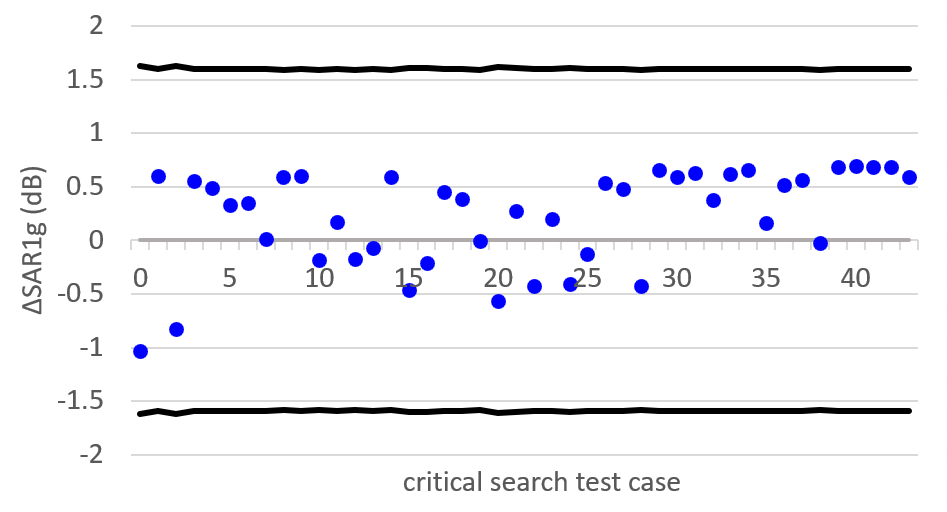}
\caption{Deviation (dots) of the SAR$_{1g}$ measured on cSAR3D from the published SAR$_{1g}$ numerical target for the 44 critical configurations identified in Table~\ref{tab:csar3}.
    All of the results are within the $\text{MPE}_j$ limits (black lines), such that the
    measurement system passes the validation.} \label{fig15}
\end{figure}

The 44 identified critical cases were measured in LAB2.  The results are shown
in Fig.~\ref{fig15}. It is observed that (\ref{eqmpe}) is met for all
measurements. Thus this cSAR3D flat phantom has passed all of the validation
criteria, and is considered to be fully validated for measurement use. Note that
this only validates the individual cSAR3D flat phantom and not the class of all
cSAR3D flat phantoms, since measurement accuracy is dependent on manufacturing
tolerances and calibration quality. Each phantom must be individually validated.

\subsection{DASY8 Validation}

The main differences in the results obtained with the DASY8 compared to cSAR3D
stem from the fact that the deterministic component of the isotropic variogram
is now small compared to the nugget (not as in the model obtained in the cSAR3D
case; see \textit{Discussion}). The obtained isotropic semivariogram and good
fit validation are summarized in Fig.~\ref{fig13a}.  As expected, given the
small sill, the NRMSE is larger than in the cSAR3D case, but it is still
comfortably within the 25~\% acceptance threshold. The residuals validation test
results (see Fig.~\ref{fig14a}) further confirm the GP model. As a result of the
prominent noise level (compared to the systematic deviations), the Shapiro-Wilk
test p-value is excellent. The slope of the QQ-plot regression is slightly below
$1$, which means that the model is on the conservative side (the estimated
interpolation errors are in average slightly larger than those obtained through
measurements), which pushes the search algorithm to use a larger initial sample
and is likely to increase the validation effort. No configuration with a
probability above $5~\%$ of exceeding the MPE value was found, which did not
come as a surprise, considering the precision of the system-under-validation. In
other words, the targeted search for critical cases was performed and no
critical points were returned.  Therefore, the evaluated DASY8 system (including
the specific probe and phantom used) is considered to be validated.

\section{Discussion}

\subsection{Revealing Device-Specific Failure Risk Heterogeneity} 

When analyzing the search algorithm performance for the cSAR3D $(x, y,
\theta)$-subcube GP model using intentionally lowered -- i.e., stricter -- MPEs,
clear clustering of the revealed configurations at risk of exceeding the MPE are
evident in Fig.~\ref{fig11} and Fig.~\ref{fig12}. The clustering around specific
spatial locations and angular orientations reveal how a combination of sensor
calibration variability and field reconstruction methodology results in a
heterogeneous distribution of the measurement error throughout the configuration
space. Despite using a validation approach that is device agnostic, the GP
model-based approach is capable of revealing such variations, and of assessing
the probability of measurement errors exceeding the tolerances.

\subsection{Behavior Comparison Between cSAR3D and DASY8}

For the DASY8 system, the deterministic component of the isotropic variogram is
almost negligible compared to the \textit{nugget}. The latter results from
thermal noise, amplifier instability, stochastic placement and other operator
inaccuracies. The cSAR3D and DASY8 system nuggets nearly have the same
magnitude, indicating that their noise levels are comparable. However, in the
cSAR3D case, the nugget only represents a small portion of the sill, indicating
that systematic (i.e., reproducible) deviations from the target value cannot be
neglected compared to the measurement noise anymore. Importantly, the GP model
approach performs correctly in both situations and remains as effective in the
presence of a high relative amount of noise as in the almost deterministic
cases. This further strengthens the confidence in the general applicability of
the developed validation approach.

The possibility of augmenting the GP model quality using manufacturer knowledge,
without compromising the trustworthiness of the validation (thanks to the
independently performed model confirmation step), could also be demonstrated for
the DASY8 system.

\subsection{Benefits of the developed methodologies}

The developed validation approach offers a range of important and valuable
features:

\begin{itemize}
\item Validation can readily be performed by an \textit{independent party}; in
    fact, different parts of the validation can be performed by different
        parties, thus maximizing trust in the validation,

\item It \textit{maximizes the likelihood} of detecting potential configurations
    that violate the MPE limit,

\item It incorporates \textit{stochastic elements}, which ensures comprehensive
    coverage over time, avoids bias, and prevents preferential calibration for
        known validation benchmark configurations,

\item It is a \textit{device-agnostic} approach where no device-specific
    knowledge is assumed, such that is ideally suited for harmonizing the
        divergent standards of scanning- and array-based systems; at the same
        time, manufacturers are free to use their knowledge to reduce the
        GP model generation effort.

\item It can be performed and repeated with a \textit{reasonable effort} 
(one day of measurements) by a suitable test-lab or the device user.

\item It can be implemented in a software tool that can \textit{easily be used} without 
any knowledge of the underlying mathematics (see the implementation accessible at \cite{bujard2}).
\end{itemize}

The proposed validation method has been demonstrated on two different SAR measurement systems.
Beyond the benefits for validation purposes, the developed methodologies also offer:
\begin{itemize}

\item an \textit{efficient approach for exploring high-dimensional parameter
    spaces} with a very small training set that furthermore is generated in a
        non-iterative manner (important in the given application context, as a
        single measurement session is desired; this is, e.g., not the case for
        the sophisticated approaches from \cite{busby} or \cite{vanbeers});

\item a \textit{search algorithm that exploits knowledge of the variogram} at
    each iteration for increased computational efficiency; this implies the
        ability to use large search populations, with no need for elaborate
        initial designs based on refined models. Increasing the search
        population allows to detect even highly local anomalies.
\end{itemize}

\subsection{Comparison to Other Approaches}

\emph{Comparison to reliability theory:} The present approach addresses a need
that is not typically covered by reliability theory  approaches (such as
\cite{azzimonti}, \cite{bect}, or \cite{dubourg}), where the determination of
the reliability of a system relies on the overall probability of failure
expressed as
\begin{align*}
P_F = \int_X I_F(x)p(x) dx
\end{align*}
for $F \subset X$ the region of failure, $I_F: X \ra \{0, 1\}$ the indicator
function of $F$, and $p$ some probability density function on $X$.  In our
application, we are not interested in the overall failure probability over a
space of events.  Instead, we want to find those $x \in X$ where the deviation
$y(x)$ has a high probability $q(x)$ of exceeding the MPE. In an optimal
scenario the failure probability would be expressible as
\begin{align*}
P_F = \sup_{x \in X} q(x)
\end{align*}
for $q(x)$ the probability of the value $y(x)$ to violate the MPE.  In this
situation the failure detection problem could be formulated as an optimization
problem:
\begin{align*}
\argmax_{x \in X} q(x).
\end{align*}
This however is not suitable for our purpose: as not all configurations $x$ are
physically measurable (only a non-dense countable subset of $X$ is), we are
interested in regions of failure rather than individual points of failure.  We
therefore need an algorithm that outputs the subset $F \subset X$ whose
individual elements have a significant probability to violate the MPE.  As
explained above, $F$ is typically disconnected and connected components can be
arbitrarily small. In general, no GP surrogate will guarantee to successfully
model all potential outliers; our geometric GP model assumptions are meant to
provide a context in which it is reasonable to assume that a large part of $F$
can be found. In the SAR validation case, data analysis has shown smoothness and
good geometricity of the underlying physics, and the present method offers a
major improvement over the published IEC 62209-3 standard which only requires a
predefined set of configurations to be validated.

\emph{Comparison to Bayesian methods:} Bayesian methods typically rely on an
acquisition function which is based on the current information provided by the
surrogate model. The search process is then guided to either improve the
surrogate model or to find more optimal function values.  In our application,
the modeler is free to use any Bayesian method to iteratively build the GP model
within Step 1 of the proposed approach, but from the point of view of a test lab
independently executing Step 3, model improvement is not allowed.  Because of
that, a separate acquisition function such as \emph{expected improvement} is not
necessary; it suffices to directly apply the search to the fixed GP model.

The iterative model improvement of Bayesian methods allows the search to neglect
the uncertainty of the the covariance function (i.e. variogram fitting) as
component of the interpolation error. Since the surrogate model is fixed and we
cannot impose an optimal level of accuracy (unlike iterative approaches), the
covariance error may lead to overestimated confidence in the predicted values
distributions.  With the proposed approach, the incorporation of the variogram
RMSE into the resulting estimated error alleviates this problem. The approach
embraces that the fitted variogram is an imperfect approximation of the
empirical one, accepts a larger fitting tolerance, and adds a safety factor to
avoid underestimation of the uncertainty.  This manifests in improved capture of
the QQ plot slope. 

Since the method goal is conservative estimation of failure risk, rather than
generation of a highly accurate model of device performance, it strikes a
different trade off between efficiency and accuracy.  While Bayesian
optimization methods typically rely on reducing uncertainty by  optimizing an
expected improvement function at each iteration, our approach accepts estimated
model uncertainty as is, and it locates the parameter value regions most at risk
of failing without reevaluating the underlying truth or adapting the error
prediction. As a result, it essentially scales linearly with the number of
trajectories. Even though the required number of trajectories is affected by the
dimensionality of the problem, it primarily depends on the global proximity of
the model to the failure thresholds.  Thus, it will remain small for systems
that are unlikely to fail, which allows the present approach to scale favorably
with dimension.

\emph{Comparison to Monte-Carlo methods:} The search algorithm should be fast
and efficient, notably in order to allow for a large initial search population
when the system performance is known to be close to the MPE.  On the other hand,
when performance is far from the MPE, critical regions can become very small, in
which case Monte-Carlo methods often suffer from the curse of dimensionality.
The $\delta$-function (which plays a role similar to the $\delta$ in the
classical analytical definitions of continuity and convergence) naturally
permits the use of particle velocities that adapt to the likelihood of exceeding
the MPE, thus accelerating evolution towards critical regions, while reducing
the risk of missing (or not converging to) small critical regions in the
configuration space.

\subsection{Applicability to Model Validation}

The elements of the presented approach can be applied for diverse forms of model
validation (e.g., in data-driven modeling and machine learning), where the GP
surrogate $Y$ is used to model an error function $e: X \ra \Rbb$ knowing the
errors at locations $S$ between the true values and the predicted values
returned by the model to be tested.  In this context, assuming a continuous and
``smooth'' response curve, the residuals validation step (notably the QQ-plot)
can be applied to locate the outliers that break the assumption of smoothness
and fail to be predicted by the GP surrogate, and the search algorithm returns
the regions of the parameter space with critical errors.  Outliers and critical
cases are then used to either validate the model or locate the regions on which
the original model is to be improved.

In the case of SAR measurement systems, the variance anisotropy is relatively
mild and the linear transformation to an isotropic version of the parameter
space does not disfigure latin hypercube samples into hyperrectangles whose
support lengths substantially vary along the different dimensions. In general,
the model uncertainty will increase along overstretched dimensions; the decrease
of information along these dimensions will increase the uncertainty of the
model, and as a result of approaching the failure threshold increase the
computational time of the search algorithm by requiring a potential significant
increase in the number of trajectories. This may affect the performance of the
method, but should not affect the outcome of the validation.  The issue can be
alleviated by applying sampling techniques that distribute points according to
an a priori estimated isotropization.

\subsection{Limitations and Extensibility}

The present approach can readily be extended in view of future evolutions of the
measurement standards, e.g., by adapting the configuration space to include
other frequencies, modulation schemes, or antennas.  Also, if future devices
prove to break certain assumptions on the model, or show the need to refine the
search, the proposed validation approach remains open for any of its components
to be generalized or improved. 

\emph{Assumptions:} The space is \textit{geometrically anisotropic} and can be
made isotropic through linear transformation; if that is not the the case,
nested model, multi-fidelity models, or multiple models valid in subspaces can
be used. In addition, a \textit{zero drift process} is assumed throughout the
space (otherwise, universal kriging is needed). \textit{Stationarity, monotonic
semivariograms, seperatability and continuity} are assumed (see above). Various
generalizations are possible at the cost of increased complexity, which
includes, replacing ordinary kriging by universal kriging, geometric anisotropy
by zonal anisotropy, and treating continuous domains with categorical variables.

\emph{Operator variability:} it is included in the noise and nugget, but is
likely operator specific (operator error might anyway trigger remeasurement if
outliers are apparent). \textit{Poor device consistency} (manufacturing or
calibration) will result in failure of confirmation step that could be remedied
by creating a specific GP model for that device.

\emph{Measurement efficiency:} The LHS nature of $S$ makes the measurement
process slow as each sample point requires a completely different measurement
configuration. However, the user can sort $S$ by antenna type (switching
antennas is the most time consuming task) to improve measurement efficiency.
This is an advantage of our approach over an iterative approach.

\emph{Other application-related aspects:} The current study focuses on
\textit{flat phantoms} and still needs to be applied to the slightly more
complicated situation in head phantoms, where the lack of translational symmetry
increases the configuration space. There are \textit{other aspects of system
validation} that are taken care of outside the GP model of this paper.
Simultaneous transmission of multiple signals (as done using 5G signals with
carrier aggregation to improve bandwidth, for example) is validated separately.
Another example that is validated separately is the validation of dynamic power
control systems of mobile phones, where the signal amplitude is adjusted over
time to keep the psSAR below the regulatory limit.

\emph{Surrogate modeling approach:} GP modeling is a wide field in which more
sophisticated approaches for model creation, validation, and exploitation exist
than the ones employed here. For instance, it might be preferable to use maximum
likelihood or cross-validation to select the covariance parameters, or the
negative log predictive density (NLPD) could complement the RMSE in the quality
assessment. In fact, other non-GP surrogate modeling approaches could be
employed, as long as they are capable of providing conservative variance
estimates (the validation step will need to be adapted if ordinary kriging is
not used).  However, in view of the purpose of this study -- namely identifying
a practical solution to the specific problem of SAR system validation and to the
more general one of efficiently and reliably validating systems with a large
configuration space in an unbiased, implementation-agnostic and independently
verifiable manner -- choices in the current study were dictated by the benefits
listed above, as well as the simplicity and ready availability of suitable
routines, and are justified by the successful application.

\section{Conclusion}

A general, robust, trustworthy, efficient, and comprehensive validation approach
has been developed that can operate agnostically of the technical implementation
of the tested device, prevents calibration that favors known validation
benchmarks, and is applicable to large configuration spaces. Its applicability,
suitability, and strength has been demonstrated through rigorous validation of
two technologically completely different SAR measurement system types, a scanning
system and an array system, and involved two different laboratories for separate
steps in the process to illustrate how the approach permits independent
verification of the validation.

The proposed approach resolves the current problem of unifying and demonstrating
equivalence of the IEC standards \cite{iec62209-1528} and \cite{iec62209-3}. At
the same time, it establishes a process by which any laboratory or user can
evaluate at any time and with reasonable effort (less than one day) that the
system performs within its reported uncertainty bounds using the commonly
available, standardized set of antennas for which target values are specified by
the standard. At the same time, the generality of the method ensures that the
set of validation antennas can readily be extended or adapted without affecting
it, as long as the exposure space is comprehensively covered.

The proposed validation approach is general (i.e., not specific to the SAR
measurement system) and has been made available as a simple software 
tool~\cite{bujard, bujard2}, so
that it can easily be followed despite the complexity of the underlying
mathematics. Overall the procedure will improve the quality, reliability, and
reproducibility of the assessment of wireless devices conformity with safety
regulations which benefits the public, government agencies and industry alike.

\section{Acknowledgment}

This research was partially supported by SEAWave (EU
HORIZON-HLTH-2021-ENVHLTH-02, Grant Agreement-101057622). We thank our
colleagues Profs. Luc Martens, and Theodore Samaras for their invaluable inputs,
Nitin Jain for performing the SAR measurements at the BNN lab. and Sabine Regel
for careful review of the manuscript. We specially thank Tim Harrington (FCC,
USA) and Gréguy Saint-Pierre (ISED, CA) for providing the perspectives of the
regulators.

\section{Conflicts of Interest}

Niels Kuster is co-founder, co-owner, and board president of SPEAG,
which manufactures SAR measurement equipment. Mark Douglas is partly employed by
SPEAG.

\section{Supporting Information}

\subsection{Geometric GP models \label{app:GP}}

This section defines geometric GP models. The background theory is well known:
it was first established in the field of geostatistics by \cite{matheron} and
detailed accounts can be found in the literature (e.g.,  \cite{chiles-delfiner}
or \cite{isaaks-srivastava}).

We let $Y = \{Y(x) : x \in X \}$ be a \emph{Gaussian process} over a convex
connected index set $X \subset \Rbb^n$; $X$ is the underlying \emph{domain} or
\emph{parameter space}. We say the process $Y$ is \emph{separable} on $X$ if it
is uniquely determined from a countable set of points (such as $X \cap \Qbb^n$),
and it is said to be (\emph{sample}) \emph{continuous} if almost all its
realizations are continuous.  More generally, a process is \emph{weakly
continuous} if it can be decomposed as a sum of a continuous process and an
uncorrelated noise component. All Gaussian processes will be assumed to be
separably weakly continuous, as well as \emph{stationary}: such a process has a
constant mean and on two points $x_1, x_2 \in X$ only depends on the difference
$h = x_2-x_1 \in \Rbb^n$, as opposed to the actual positions of these two points
in $X$. If the process only depends on the euclidean distance $|h|$, it is said
to be \emph{isotropic}, otherwise it is \emph{anisotropic}. The process is
\emph{geometric} if there exists an invertible linear map $\iota : X \ra
\iota(X)$ in $GL_n(\Rbb)$ such that the process $Y_\iota = Y \circ \iota^{-1}$
on $\iota(X)$ is isotropic.  Clearly, an isotropic process is geometric.

The \emph{semivariogram} $\gamma : \Rbb^{2n} \ra \Rbb$
can be defined via
\begin{align}
    2 \gamma(x_1, x_2) = \Var \left(Y(x_1)-Y(x_2)\right). \label{eq2}
\end{align}
The stationary nature of $Y$ implies that $\gamma$ is also defined as a univariate
function $\gamma : \Rbb^n \ra \Rbb$ by $\gamma(h) = \gamma(0, h)$ via the
relation $\gamma(x_1, x_2) = \gamma(x_2 - x_1)$. The corresponding
\emph{variogram} is $2 \gamma$.  

If the process is furthermore isotropic, $\gamma$ can be defined on $\Rbb$
by $\gamma(|h|) = \gamma(h)$, in which case $\gamma$ is called \emph{isotropic}.
More generally for a subvector space $V$ of $\Rbb^n$, a semivariogram $\gamma$
is said to be \emph{directional} on $X$ along $V$ if $\gamma$ is isotropic on $X
\cap V$.  Obviously, an isotropic semivariogram is directional, and any
directional semivariogram $\gamma'$ can be defined on $\Rbb$ by setting
$\gamma'(|h|) = \gamma(h)$. In this study, $\gamma$ is assumed to be monotonic on
$|h|$ and to be directional on some subspace of $X$.  As the expectation of a
square, $\gamma(x_1, x_2) \geq 0$ for all $x_1, x_2$.  At lag $h = 0$,
$\gamma(0) = \gamma(x, x) = \E\left[(Y(x)-Y(x))^2\right] = 0$, such that
$\gamma$ is always zero at the origin.  Nevertheless, the limit
\[
    n_\gamma = lim_{h \ra 0} \gamma(h)
\] 
can be non-zero and is called the \emph{nugget}. When $n_\gamma>0$, it describes
the height of the discontinuous jump at the origin. In case $X$ has no spatial
dependence the variogram is the constant $\Var(Y(x))$ everywhere except at the
origin, where it is zero.  As explained in \cite{chiles-delfiner} paragraph
2.3.1 it can be shown that a stationary separable Gaussian process is continuous
if and only if it is \emph{mean square continuous} in the sense that
\[
    \lim_{|h| \ra 0} \E[|Y(x+h) - Y(x)|^2] = 0 \q \text{for all } x \in X,
\]
and this is true if and only if $n_\gamma = 0$.  When $n_\gamma \neq 0$ it is
still assumed that $Y$ is weakly continuous, so that the origin is the only
point of discontinuity in the variogram.

In the directional case, the limit 
\[
    s_\gamma = \lim_{|h|\to \infty} \gamma(h) 
\]
exists and is called the \emph{sill} of the semivariogram $\gamma$. The lag
$r_\gamma$ at which $s_\gamma - \gamma(h)$ becomes negligible is called the
\emph{range}: when $s_\gamma = \gamma(h)$ for some $h$, then $r_\gamma$ is the
infimum of all such $h$; while for models with an asymptotic sill, $r_\gamma$ is
conventionally taken to be the distance at which the semivariance first reaches
95~\% of the sill, in which case it is referred to as the \emph{effective range}.

Since in practice it is impossible to sample everywhere, the interval $[0,
\infty)$ is binned and for a finite set of known sample $S = \{y_i\}$, the
function $\gamma$ is estimated based on the empirical variogram $\hat{\gamma}$
defined by
\begin{align}
    \hat{\gamma}(h)
    :=
    \frac {1}{2 N(B_h)}
    \sum_{|y_i - y_j| \in B_h} (y_{i} -y_{j})^2, \label{eq3}
\end{align}
for $B_h$ the bin that contains $h$ and $N(B_h)$ the cardinality of the set
$\{|y_i - y_j| \in B_h\}$; this estimator is also refered to as the
\emph{Matheron semivariance estimator}. Of course the binning, the size and
evenness of the sample all play major roles in the quality of the resulting
$\hat{\gamma}$.  A theoretical variogram model is then fit to the empirical
values -- we say that $\gamma$ \emph{fits} $Y$ on $S$. Three of the most common
models are:
\begin{itemize}
    \item the \emph{Exponential} model
    \begin{align}
        \gamma(h) &= n_\gamma + s_\gamma 
        \left( 1 - e^{- \frac{3h}{r_\gamma}} \right),
        \label{eq4}
    \end{align}

    \item the \emph{Gaussian} model
    \begin{align}
        \gamma(h) = n_\gamma + s_\gamma 
        \left( 1 - e^{- \frac{4h^2}{r_\gamma^2}} \right),
        \label{eq5}
    \end{align}

    \item the \emph{Spherical} model
    \begin{align}
        \gamma(h) = n_\gamma + \frac{s_\gamma}{2 r_\gamma}
        \left( 3h-h^3 \right).
        \label{eq6}
    \end{align}
\end{itemize}

We can now define a \emph{geometric GP model} to be a commutative diagram
\begin{align}
\xymatrixrowsep{3pc}
\xymatrixcolsep{4pc}
\xymatrix
{
    S \ar[dr]_{y} \ar[r]^{\text{inc}} & X \times \Omega \ar[d]_{Y} \ar[r]^{\iota
    \times \text{id}}_{\cong} &
    \iota(X) \times \Omega \ar[dl]^{Y_\iota} \\
    & \Rbb
}
\label{eq7}
\end{align}
endowed with a monotonic isotropic semivariogram $\gamma: \Rbb^+ \rightarrow \Rbb^+$ that fits
$Y_\iota$ on $S$, where:
\begin{itemize}
    \item $S = \{x_i\}$ is a finite subset of $X \subset \Rbb^n$; 
    \item
        $Y = \{y : X \rightarrow \Rbb\}$ is a geometric, separable, stationary, and 
        weakly continuous Gaussian process with probability space $\Omega$; 
    \item $Y_\iota$ is isotropic with $\iota$ an invertible linear map in $GL_n(\Rbb)$.
\end{itemize}
The commutativity of the diagram implies the process is fully known on $S$, with
\[
    Y(x_i, \omega) = y(x_i) =: y_i
\]
for all $x_i \in S$, $\omega \in \Omega$, $y \in Y$.
Writing $\bar{S} = \{ (x, Y(x)) : x \in S) \} \subset \Rbb^{n} \times \Rbb$, 
a geometric GP model is characterized
by the triple $(\bar{S}, \iota, \gamma)$ which can be used to fully represent
the model in a software implementation. Under the geometric GP model assumptions,
$\bar{S}$ is sufficient to build a full model. The approach is as follows: 
\begin{enumerate}
    \item Given $\bar{S}$, one can build $\iota$ by probing spatial dependencies
        along all possible ($1$-dimensional) directions in the data space. Under
        the geometric assumption of a GP model an iso-variance contour forms an
        $(n-1)$-dimensional ellipsoid in $X$. This ellipsoid can be mapped to an
        $(n-1)$-sphere via a series of rotations and axis rescalings. The
        composition of these invertible linear maps on $\Rbb^n$ form $\iota :
        \Rbb^n \ra \Rbb^n$ which can be represented as a matrix in the canonical
        base of the data space.

    \item Given $\bar{S}$ and $\iota$, one can compute an empirical
        semivariogram $\hat{\gamma}$ on the isotropic space $\iota(X)$, on which
        a semivariogram $\gamma$ can be fitted. Most often, the model and
        binning are set a priori -- they depend on the application field --, while
        the range, sill and nugget are the result of a fitting algorithm. 
\end{enumerate}

A geometric GP model can then be used for spatial inference at any unobserved
location via ordinary kriging. The unknown value $Y(x)$ and the (unbiased)
ordinary kriging estimator $\hat{Y}(x)$ are interpreted as random variables
located at $x$. The linear combination $\epsilon(x) := \hat{Y}(x) - Y(x)$ is a
Gaussian random variable with zero mean, and for $\sigma^2(x)$ the kriging
estimator variance, the variable
\begin{align}
    \frac{\epsilon(x)}{\sigma(x)} 
    = \frac{\hat{Y}(x) - Y(x)}{\sigma(x)}  
    \sim \Ncal(0, 1)
    \label{eq16}
\end{align}
follows the standard normal distribution. 

\subsection{The Delta Function\label{app:deltameasure}}

Let $T \in \Rbb$ be a fixed threshold and $\Mcal = (\bar{S}, \iota, \gamma)$ a
geometric GP model.  For $\{Y(x) : X \ra \Rbb \}$ the underlying geometric Gaussian
stationary process over $X \subset \Rbb^n$ known on $S = \{x_i\}_{i = 1}^k$, and
for $x_0$ a point in $X$, a measure needs to be established of how far away from
$x_0$ the next candidate point $x$ must be to have a reasonable likelihood of
$Y(x)$ \emph{crossing} the threshold $T$ in the sense that:
\[
    \left(Y(x) - T\right) \left(Y(x_0) - T\right) < 0.
\]
The probability $p$ that $Y(x) < T$ knowing that $y_0 = Y(x_0)$ for a neighboring point
$x_0$ is
\begin{align*}
    p &= P(Y(x) < T\ |\ Y(x_0) = y_0) \\
      &= P(Y(x)-Y(x_0) < T-y_0\ |\ Y(x_0) = y_0) \\
      &= \Phi\left(\frac{T-y_0}{\sigma}\right),
\end{align*}
where $\sigma^2$ is the variance of the zero-mean Gaussian distributed variable
$Y(x)-Y(x_0)$. In this case, the probability that $Y(x) > T$ knowing $y_0 =
Y(x_0)$ is $1-p$. Since by definition $\sigma^2 = 2\gamma$, assuming $p <
\frac{1}{2}$,
\begin{align*}
    &\Phi\left(\frac{T-y_0}{\sigma}\right) = p \\ 
    &\q\Lra\q T-y_0 
        = \sigma \Phi^{-1}(p)
        = \sqrt{2\gamma(h)} \Phi^{-1}(p) \\
    &\q\Lra\q \gamma(h) 
        = \frac{1}{2} \left( \frac{T-y_0}{\Phi^{-1}(p)} \right)^2,
\end{align*}
for the lag $h = x-x_0 \in \Rbb^n$. Along a given direction, $\gamma$ is
invertible on the interval $(0, r_\gamma]$ with $s_\gamma := \gamma(r_\gamma)$
for $r_\gamma$ the (effective) range of $\gamma$. Then the
continuous monotonic function $\delta_p : \Rbb_+ \ra \Rbb_+$ can be defined as 
\begin{align}
    \delta_p(l) = 
    \begin{cases}
        0 &l \leq g_p(n_\gamma),\\
        \gamma^{-1} \left( \frac{l^2}{2\Phi^{-2}(p)} \right) &g_p(n_\gamma) < l < g_p(s_\gamma),\\
        r_\gamma &l \geq g_p(s_\gamma),
    \end{cases}
\end{align}
where
\begin{align}
    \q g_p(d) = \sqrt{2 d} |\Phi^{-1}(p)|, \qq 0 < p < \frac{1}{2},
\end{align}
for $n_\gamma$ the nugget of $\gamma$. Knowing the values $T$ and $y_0$ one can
choose a parameter $p$ and use $\delta_p(T-y_0)$ to find the closest points to
$x_0$ with a probability $\geq p$ to cross the threshold $T$. By construction
the function $\delta_p$ encodes all the covariance characteristics carried by
the variogram:
\begin{itemize}
\item Parameter $p$ expresses the sensitivity of the model in the assessment of
        whether or not the true value $y(x)$ is considered to have crossed $T$.

\item Models typically assume $n_\gamma$ to be zero. However, when $|T-y_0|$ is
    smaller than a non negligible $g_p(n_\gamma) > 0$, the expected value of
        $Y(x)$ is so close to $T$ that $\delta_p(l)$ is zero, which expresses
        the fact that it cannot be predicted on what side of $T$ the true value
        $y(x)$ is likely to be.   

\item As $y_0$ approaches $T$, $\delta_p(l)$ approaches zero in accordance with
    $\gamma$.

\item When $|T-y_0|$ is larger than $g_p(s_\gamma)$, $x$ is too far away from
    $x_0$ for the condition $Y(x_0) = y_0$ to have an influence on the outcome
        of $Y(x)$.
\end{itemize}

The function $\delta_p$
can only be effectively incorporated into an algorithm, if it can be efficiently
evaluated at any $x$. This depends on the ability to efficiently evaluate
the inverse function of the variogram $\gamma$ on the interval $(n_\gamma,
r_\gamma)$ defined by the nugget and range parameters. Even though computing the
inverse of a function can be computationally intensive (provided it even
exists), the most commonly used variogram models (see (\ref{eq4}), (\ref{eq5})
and (\ref{eq6})), including the one used here, happen to be analytically
invertible.

\subsection{Size of Initial Search Population\label{app:initalpopulation}}

How can it be ensured that $S_0$ is suitable for the initiation of algorithm
\ref{alg1}? A method is needed to determine the sample size required for the
search algorithm to either locate these out of bound elements, or -- if none are
found -- to establish confidence that there are none. It must be applicable even
in those cases where the likelihood of crossing $T_-, T_+$ is low (i.e., the
$T_\ast$ are far from the known values $Y(S_0)$).

Assume that $T_- < Y(x) < T_+$ for every $x$ in $S_\ast$ (i.e., the device is
valid). For $x_0 \in S_\ast$ and $T \in \{T_-, T_+\}$, it is known by definition
that for any continuous realization $y$ of $Y$ and for any $l = |y(x_0)-T|$,
there is a $\delta > 0$ such that none of the elements in the open hyperball
$B(x_0, \delta) \subset X$ with center $x_0$ and radius $\delta$ will cross $T$.
Determining the supremum $\delta_{\text{sup}}(y)$ of all these $\delta$'s for
each continuous $y$, and setting $\delta_{\text{inf}}$ to be the infimum of all
$\delta_{\text{sup}}(y)$ would give a neighborhood around $x_0$ in which an
element $x$ is known not to cross one of the thresholds.  The problem is that
without further conditions on a realization $Y$ it can reach arbitrarily large
values in any given open neighborhood of $x_0$, even if it is continuous -- in
other words $\delta_{\text{inf}} = 0$.  It is therefore impossible to guarantee
with a finite sample in $X$ and a finite searching algorithm that all out of
bound elements will be found. However, by the weakly continuous nature of the
underlying process, and by the definition of $\delta_p$, it is possible to
establish a condition with probability $0 < p < \frac{1}{2}$ in which
$\delta_{\text{inf}}$ reaches $\delta_p(l)$: For a finite sample $S_\ast \subset
X$ in an isotropic space $X$ with semivariogram $\gamma$, an element $x_0 \in
S_\ast$, and a threshold $T_\ast \in \Rbb$, the largest $\delta > 0$ with a
probability of at most $p$ for elements of $B(x_0, \delta)$ to cross $T_\ast$
(knowing $\gamma$) is $\delta = \delta_p(|Y(x_0)-T_\ast|)$.  One can be
confident, with probability $q = 1-p$ for small $p > 0$, that all the points in
the open subset of radius $\delta_p(l)$ around $x_0$ will not cross $T$.
Ideally, the situation should be reevaluated by performing additional
measurements at points in $x \in X$ that are distant by more than $\delta_p(l)$
from $x_0$. In practice, because algorithm \ref{alg1} searches along a single
dimension at a time, it is sufficient to consider the $2\cdot \dim(X)$ points on
the hypersphere $S(x_0, \delta) = \partial B(x_0, \delta)$ that are located
along directions $\{\overrightarrow{e_i}\}$.

The closer $y(x)$ is to $T_\ast$, the smaller $\delta$ becomes; therefore, for
each $x \in S_\ast$ one is only interested in the threshold $T$ that is the
closest to $y(x)$.  Assuming that the values of $S_\ast$ are evenly distributed
over $X$, let
\[
    L_\ast = \{ \min(|y(x)-T_-|, |y(x)-T_+|)\ :\ x \in S_\ast \}
\]
and define $\bar{l}$ to be the sample mean of $L_\ast$. Then, for $X_i$ (the
projection of $X$ on its $i$th dimension; $1 \leq i \leq n = \dim(X)$) the
positive integer
\begin{align}
    \nu_p = \prod_{i = 1}^{n}
    \left\lceil\frac{\sup(X_i)-\inf(X_i)}{\delta_p(\bar{l})}\right\rceil
\end{align}
is suggested as an estimate of the required sample size. Here, $\nu_p$ is the
number of points of the coarsest discretization of $X$ resulting from uniform
discretizations of all $X_i$ with step interval at most $\delta_p(\bar{l})$. In
practice, a uniformly distributed sample of size $\nu_p$ over $X$ that
incorporates a random element is desirable, such as a $n$-dimensional latin
hypercube sample that maximizes the minimal distance between sample points.
Clearly, $\nu_p$ must and does increase when the required confidence $q =
1-p$ increases. As expected
    \[
        \lim_{q \ra 1} \nu_p = \lim_{p \ra 0} \nu_p = \infty.
    \]

From the above, it follows that the proposed approach permits $p$ to be set
based on effort-reliability-balance considerations and ensures that the
algorithm detects threshold violations with a user-defined sensitivity level.


\begin{IEEEbiographynophoto}{Cédric Bujard} 
was born in Lausanne, Switzerland in 1978. In 2001 he received the BA in Music
from Berklee, Boston, in 2006 the MSc in Mathematics from the Swiss
Federal Institute of Technology Lausanne (EPFL), Switzerland, and in 2012 the PhD
in Mathematics from the University of Strasbourg, France. 

From 2006 to 2007, he worked as Assistant in EPFL, then until 2012 as Lecturer
in EPFL. Since 2012 he has been working as a Mathematical Modeler, Data
Modeler, Data Scientist, and Software Engineer; working since 2021 at the IT’IS
Foundation in Zurich, Switzerland.  
\end{IEEEbiographynophoto}

\begin{IEEEbiographynophoto}{Esra Neufeld} 
received an MSc in Interdisciplinary Sciences in 2004 and a PhD in
Electrical Engineering in 2008, both from the Swiss Federal Institute of
Technology (ETH) Zurich, Switzerland. 

He has been Head of Computational Life Sciences at the IT'IS Foundation in
Zurich since 2008 and its Associate Director since 2018. He has also been
Chief Science Officer of ZMT Zurich MedTech AG since 2009 and co-founded TI
Solutions AG in 2019.  Dr. Neufeld has published over 80 journal articles and book chapters. His
research covers the field of computational life-sciences, with a particular
interest in understanding the biological/physiological responses of living
tissues and organisms to physical exposure and in developing and employing
in silico methods for the development of safe and effective therapies and
devices.  He is involved in standardization activities and frequently
advises regulatory authorities and government agencies on in silico methods.
\end{IEEEbiographynophoto}

\begin{IEEEbiographynophoto}{Mark Douglas} 
(Senior Member, IEEE) was born in Vancouver, Canada in 1967. He received the
B.Eng. degree from University of Victoria, Canada in 1990, the M.Sc. degree
from University of Calgary, Canada in 1993, and Ph.D. degree from University
of Victoria in 1998, all in electrical engineering. 

He was with Ericsson EMF Research Group
from 1998 to 2002 and Motorola Electromagnetic Energy Research Laboratory
from 2002 to 2009. Since 2009 he has been a Project Leader at the IT’IS
Foundation in Zurich, Switzerland. 

Dr. Douglas's research work in electromagnetic dosimetry has resulted in 5
patents and over 100 papers for scientific conferences and peer reviewed
journals. He serves as the co-chair of IEEE International Committee for
Electromagnetic Safety (ICES) Technical Committee 34.
\end{IEEEbiographynophoto}

\begin{IEEEbiographynophoto}{Joe Wiart} 
(Senior Member, IEEE) received the Engineer of telecommunication
diploma in 1992, the Ph.D. degree, in 1995, and the HDR in 2015. 

Since 2015, he has been the holder of the Chair C2M "Caractérisation,
modélisation et maîtrise of the Institut Mines Telecom" at Telecom Paris.
His works gave rise to more than 150 publications in journal. His research
interests are experimental, numerical methods, machine learning, and
statistic applied in electromagnetism, dosimetry and exposure monitoring. 

Dr. Wiart has been an Emeritus Member of The Society of Electrical Engineers
(SEE), since 2008. He is also the Chairman of the TC106x of the European
Committee for Electrotechnical Standardization (CENELEC) in charge of EMF
exposure standards. He is also the past Chairman of the International Union
of Radio Science (URSI) commission k and has been the Chairman of the French
chapter of URSI.
\end{IEEEbiographynophoto}

\begin{IEEEbiographynophoto}{Niels Kuster} 
(Fellow Member, IEEE) (1957) earned his MS and PhD degrees in Electrical
Engineering from the Swiss Federal Institute of Technology (ETH) in Zurich.
He joined the academic staff of the Department of Electrical Engineering of
ETH as Assistant Professor in 1993 and was awarded the title of Professor in
2001.  During his career, Niels Kuster has also held invited professorships
at the Electromagnetics Laboratory of Motorola Inc. in Florida, USA, in
1992, and at the Metropolitan University of Tokyo, Japan, in 1998.

In 1999, he established the Foundation for Research on Information
Technologies in Society (IT’IS), where he continues today as the founding
Director. He also (co-) founded several spin-off companies, including Schmid
\& Partner Engineering AG (SPEAG) in 1994, MaxWave AG in 1998, ZMT Zurich
MedTech AG in 2006, BNNSPEAG in 2012, and TI Solutions AG in 2019. 

Prof. Dr. Kuster's main research interests are focused on (i) experimental and
computational electromagnetics from subHz to 300 GHz in complex
environments, (ii) computational life sciences, (iii) virtual human and
animal anatomical models functionalized with dynamic tissue models and (iv)
wireless and medical applications such as minimization/optimization of
body-mounted transmitters/electrode configurations, neurostimulation,
implant safety, in silico clinical trials, etc. He has published
over 250 peer-reviewed publications on measurement techniques, computational
electromagnetics, dosimetry, exposure assessment, and bioexperimentation. He
is a member of several standardization bodies and has been consulted by
government agencies around the globe on the safety of mobile communications.
He has been honored with several scientific awards and has served on the
boards of various scientific societies and journals.
\end{IEEEbiographynophoto}

\EOD


\begin{thebibliography}{00}

\bibitem{iec62209-1528}
{IEC/IEEE} 62209-1528.
\newblock {{M}easurement Procedure for the Assessment of Specific Absorption Rate of Human Exposure to Radio Frequency Fields From Hand-Held and Body-Worn Wireless Communication Devices—Human Models, Instrumentation and Procedures (Frequency Range of 4 {MHz} to 10 {GHz})}.
\newblock {International Electrotechnical Commission Geneva, Switzerland, 2020.}

\bibitem{iec62209-3}
International Electrotechnical Commission.
\newblock {{IEC} 62209-3 {M}easurement procedure for the assessment of specific absorption rate of human exposure to radio frequency fields from hand-held and body-mounted wireless communication devices – Part 3: Vector measurement-based systems (Frequency range of 600 {MHz} to 6 {GHz})}.
\newblock {Geneva, Switzerland, 2019}.

\bibitem{bujard}
Bujard C, Neufeld E, Douglas M, Wiart J, Kuster N. 
\newblock{A Gaussian process model based approach for validation of multi-variable measurement systems: application to SAR measurement systems [Computer software]}. 
\newblock {https://github.com/ITISFoundation/publication-IEC62209}

\bibitem{bujard2}
Bujard C, Neufeld E, Douglas M, Kuster N. 
\newblock{SAR System Validation Procedure}.
\newblock{[User website] http://sarvalidation.site}

\bibitem{ecRF}
Council of the European Union.
\newblock {1999/519/EC: Council Recommendation of 12 July 1999 on the limitation of exposure of the general public to electromagnetic fields (0 Hz to 300 GHz)}.
\newblock {1999 Jul; 199:59-70, http://data.europa.eu/eli/reco/1999/519/oj}.

\bibitem{cenelec50360-2017}
European Committee for Electrotechnical Standardization.
\newblock {{CENELEC EN} 50360:2017/A1:2023 -- {P}roduct standard to demonstrate the compliance of wireless communication devices, with the basic restrictions and exposure limit values related to human exposure to electromagnetic fields in the frequency range from 300 {MHz} to 6 {GHz}: devices used next to the ear}.
\newblock {Brussels, Belgium, 2023.}

\bibitem{cenelec50566-2017}
European Committee for Electrotechnical Standardization.
\newblock {{CENELEC EN} 50566:2017/A1:2023 -- Product standard to demonstrate the compliance of wireless communication devices with the basic restrictions and exposure limit values related to human exposure to electromagnetic fields in the frequency range from 30 MHz to 6 GHz: hand-held and body mounted devices in close proximity to the human body}.
\newblock {Brussels, Belgium, 2023}.

\bibitem{icnirp}
{International Commission on Non-Ionizing Radiation Protection}.
\newblock {Guidelines for limiting exposure to electromagnetic fields (100 {kHz} to 300 {GHz)}}.
\newblock {Health Physics, 2020;118(5):483--524.}

\bibitem{ieeec951}
IEEE.
\newblock {IEEE Std C95.1-1999 -- {IEEE} Standard for Safety Levels with Respect to Human Exposure to Radio Frequency Electromagnetic Fields, 3 {kHz} to 300 {GHz}}.
\newblock {Piscataway, NJ, USA, 1999}.

\bibitem{pokovic2000}
Pokovic K, Schmid T, Kuster N. 
\newblock {Millimeter-resolution E-field probe for isotropic measurement in lossy media between 100 MHz and 20 GHz.} 
\newblock {IEEE Transactions on Instrumentation and Measurement. 2000 Aug;49(4):873-8.}

\bibitem{csar3d2011}
Douglas MG, Kuster N. 
\newblock{Novel fast SAR methods for compliance testing of wireless devices.}
\newblock{International Symposium on Electromagnetic Compatibility, Tokyo. IEEE, 2014.}

\bibitem{schmid1996}
Schmid T, Egger O, Kuster N.
\newblock{Automated E-field scanning system for dosimetric assessments}.
\newblock {IEEE Transactions on Microwave Theory and Techniques. 1996;44(1):105--113.}

\bibitem{meyer2015}
Meyer R, Kühn S, Pokovic K, Bomholt F, Kuster, N.
\newblock {Novel Sensor Model Calibration Method for Resistively Loaded Diode Detectors}. 
\newblock {IEEE Transactions on Electromagnetic Compatibility. 2015;57(6):1345--1353.}

\bibitem{isedcomparison}
{Innovation, Science and Economic Development Canada}.
\newblock {Inter-laboratory fast SAR measurement campaign: overview of findings and recommendations}.
\newblock {Report to IEC Project Team 62209-3, 2017}.

\bibitem{watanabe19}
Nagaoka, T, Wake K, Watanabe, S.
\newblock {Comparison of SARs measured by vector probe array-based SAR measurement systems using commercially available smartphones}.
\newblock {Bioelectromagnetics Society Annual Meeting. Montpellier, France, 2019}.

\bibitem{jrc}
Chountala C, Cerutti I, Ferragut J, Chareau JM, Bishop J, Viaud P, 
\newblock {Standards for the Measurement of the Specific Absorption Rate -- A study on IEC 62209-1/2 and 62209-3 (full SAR measurement methods using robotic scanning systems and scalar probes and advanced methods using vector probes in an array)}.
\newblock{European Commission Joint Research Centre Technical Report, July 2023}.

\bibitem{chiles-delfiner}
Chilès, J-P, Delfiner, P.
\newblock {Geostatistics: Modeling Spatial Uncertainty, Second Edition}.
\newblock {Wiley Series in Probability and Statistics. 2012.}

\bibitem{isaaks-srivastava}
Isaaks E, Srivastava RM.
\newblock {An Introduction to Applied Geostatistics}.
\newblock {Oxford University Press, 1989, ISBN 0-19-505012-6}.

\bibitem{malicke}
Mälicke M.
\newblock {SciKit-GStat 1.0: A SciPy flavoured geostatistical variogram estimation toolbox written in Python}.
\newblock {Geoscientific Model Development. 2022; 15:2505-–2532}.

\bibitem{roustant2012dicekriging}
Roustant O, Ginsbourger D, Deville Y.
\newblock {DiceKriging, DiceOptim: Two R packages for the analysis of computer experiments by kriging-based metamodeling and optimization}.
\newblock {Journal of statistical software. 2012; 51:1--55}

\bibitem{mckay}
McKay MD, Beckman RJ, Conover, WJ.
\newblock {A comparison of three methods for selecting values of input variables in the analysis of output from a computer code}.
\newblock {Technometrics. American Statistical Association. 1979; 21(2):239--245.}

\bibitem{park}
Park JS.
\newblock{Optimal Latin-hypercube designs for computer experiments}.
\newblock{Journal of Statistical Planning and Inference. 1994; 39(1):95--111}.

\bibitem{shapiro-wilk}
Shapiro SS, Wilk MB.
\newblock {An analysis of variance test for normality (complete samples)}.
\newblock {Biometrika. 1965; 52(3-4):591–-611}.

\bibitem{gum}
Joint Committee for Guides in Metrology,
\newblock{Evaluation of measurement data — Guide to the expression of uncertainty in measurement}.
\newblock{JCGM 100:2008, September 2008}.

\bibitem{sanchez2008}
S{\'a}nchez-Hern{\'a}ndez DA.
\newblock {Multiband integrated antennas for 4G terminals}.
\newblock {Artech House. 2008.}

\bibitem{baudin2013pydoe}
Baudin M, Christopoulou M, Colette Y, Martinez, JM.
\newblock {pyDOE: The experimental design package for python}.
\newblock {Available at https://pythonhosted.org/pyDOE/\#credits}.

\bibitem{busby}
Busby, D. 
\newblock {Hierarchical adaptive experimental design for Gaussian process emulators}. 
\newblock {Reliability Engineering \& System Safety, Vol. 94(7), 2009.}

\bibitem{vanbeers}
Van Beers et al. 
\newblock{Customized sequential designs for random simulation experiments: Kriging metamodeling and bootstrapping.} 
\newblock{European journal of operational research 186.3: 1099-1113. 2008.}

\bibitem{azzimonti}
Azzimonti, D., et al. 
\newblock{Adaptive design of experiments for conservative estimation of excursion sets.} 
\newblock{Technometrics 63.1: 13-26. 2021}

\bibitem{bect}
Bect, J, Ginsbourger, D, Li, L, Picheny, V, Vazquez, E. 
\newblock {Sequential design of computer experiments for the estimation of a probability of failure}. 
\newblock {Statistics and Computing, 22(3), 773-793. 2012.}

\bibitem{dubourg}
Dubourg, V, Sudret, B, Deheeger, F. 
\newblock {Metamodel-based importance sampling for structural reliability analysis}. 
\newblock {Probabilistic Engineering Mechanics, 33, 47-57. 2013.}

\bibitem{matheron}
Matheron, G.
\newblock {Principles of geostatistics}.
\newblock {Economic Geology. 1963;58:1246-–1266}.

\end{thebibliography}
\end{document}